\documentclass[a4paper]{article}

\usepackage{longtable}
\usepackage{amssymb}

\def\va{{a}}
\def\vb{{b}}
\def\vc{{c}}
\def\vd{{d}}
\def\ve{{e}}
\def\vf{{f}}
\def\vg{{g}}
\def\vh{{h}}
\def\sa{{\alpha}}
\def\sb{{\beta}}

\def\ca{{\dot\alpha}}
\def\cb{{\dot\beta}}
\def\cc{{\dot\gamma}}

\def\sstab{{\underline{(\alpha\beta)}}}
\def\sstcd{{\underline{(\gamma\delta)}}}
\def\sctab{{\underline{(\dot\alpha\dot\beta)}}}
\def\sctcd{{\underline{(\dot\gamma\dot\delta)}}}
\def\ua{{i}}
\def\ub{{j}}
\def\uc{{k}}

\def\uA{{I}}
\def\uB{{J}}

\def\Ua{{A}}
\def\Ub{{B}}

\def\efa{{\tilde\mathcal A}}
\def\efb{{\tilde\mathcal B}}
\def\efc{{\tilde\mathcal C}}
\def\efd{{\tilde\mathcal D}}
\def\eaa{{\mathcal A}}
\def\eab{{\mathcal B}}

\def\ma{{\tt a}}
\def\mb{{\tt b}}
\def\mc{{\tt c}}
\def\md{{\tt d}}
\def\mm{{\tt m}}
\def\mn{{\tt n}}
\def\mp{{\tt p}}
\def\mq{{\tt q}}
\def\mi{{\tt i}}
\def\mj{{\tt j}}

\begin{document}

\begin{center}
{\bf\Large Fourteen new stationary points in the 
scalar potential of $SO(8)$-gauged $\mathcal{N}=8$,
$D=4$ supergravity}

\bigbreak

{\bf Thomas Fischbacher\\}
\smallbreak
{\em University of Southampton\\
  School of Engineering Sciences\\
  Highfield Campus\\
  University Road, SO17 1BJ Southampton, United Kingdom\\}
{\small {\tt t.fischbacher@soton.ac.uk}}

\end{center}

\begin{abstract}
  \noindent The list of six previously known nontrivial stationary
  points in the scalar potential of $\mathcal{N}=8$, $D=4$
  supergravity with gauge group~$SO(8)$ is extended by fourteen new
  entries, whose properties have been obtained numerically using the
  sensitivity backpropagation technique. Eight of the new solutions
  break the gauge group completely, while three have a residual
  symmetry of~$U(1)$. Three further ones break the gauge group
  to~$U(1)\times U(1)$. While the approximate numerical data are
  somewhat inconclusive, there is evidence that one of these may have
  a residual $\mathcal{N}=1$ supersymmetry, hence correspond to a
  stable vacuum. It must be pointed out that this list of new
  solutions most likely is not exhaustive.
\end{abstract}

\section{Introduction}

\noindent The maximally supersymmetric field theory in four dimensions
-- $\mathcal{N}=8$ supergravity~\cite{Cremmer:1979up} -- that is
obtainable by dimensional reduction of eleven-dimensional
supergravity~\cite{Cremmer:1978km} is quite remarkable in many ways:
The high degree of supersymmetry unifies the entire particle spectrum
in a single supermultiplet, the model features an `unexpected' hidden
global exceptional~$E_{7(7)}$ symmetry (see e.g.~\cite{Cremmer:1997ct,
  Cremmer:1998px}), and up to the possibility of a deformation that
promotes the vector fields to nonabelian gauge bosons~\cite{de
  Wit:1982ig, Hull:1984vg, Hull:1984qz,de Wit:2007mt}, its structure
seems to be uniquely determined. As the model with gauge
group\footnote{This work uses the designations of nonabelian groups
  that are customary in supergravity throughout, even in cases where,
  strictly speaking, it would be more appropriate to denote them as a
  different quotient of the universal covering group}~$SO(8)$ arises
from the compactification of~$D=11$ supergravity on a highly symmetric
space -- the seven-sphere~$S^7$ -- it occupies a prominent place in
the family of supersymmetric field theory models, and as such it is
not surprising to see that detailed information about its properties
turned out to have a number of interesting
applications~\cite{Ahn:2009et, Benna:2008zy, Distler:1998gb,
  Duff:2002rw, Nicolai:1985hs}. Most of these are, of course, related
to the AdS/CFT correspondence~\cite{Maldacena:1997re}, as the vacua of
$SO(8)$-gauged~$\mathcal{N}=8$ supergravity correspond to stationary
points in the scalar potential with negative (Planck-scale)
cosmological constant, hence an Anti de-Sitter background geometry.

Considering the recent discovery that~$\mathcal{N}=8$ supergravity
seems to be much more well-behaved at high loop orders than originally
thought~\cite{Bern:2006kd, Bern:2009kd}, and has an especially simple
$S$-matrix, it might even be justified to claim it to be ``the
simplest (interacting) quantum field theory'' (implicitly: in four
spacetime dimensions, and with a proper Yang-Mills kinetic 
term)~\cite{ArkaniHamed:2008gz}.

Given the prominence of four-dimensional $SO(8)$-gauged
$\mathcal{N}=8$ supergravity, it is somewhat remarkable that little
progress has been made on its vacuum structure during the last~25
years: While a detailed analysis by N.~Warner in~1983 managed to
produce an exhaustive list of all stationary points in the scalar
potential with residual (unbroken) gauge symmetry of at
least~$SU(3)$~\cite{Warner:1983vz}, and a further one that
breaks~$SO(8)$ to~$SO(3)\times SO(3)$,~\cite{Warner:1983du}, no new
candidates for vacua of this model could be obtained despite repeated
attempts (see e.g.~\cite{Fischbacher:2003rb, Ahn:2009as}). The
underlying reason for this is easy to understand: The $70$~scalars of
the model can be thought of as parametrizing the coset
space~$E_{7(7)}/SU(8)$, so an analytic investigation of the full
potential would seem to require an ``Euler angle'' type
parametrization of that space. Even when taking the $SO(8)$~symmetry
of the potential into account that reduces the number of degrees of
freedom to~$70-28=42$, the task of determining analytic expressions
for such a parametrization of part of~$E_7$ in terms of
complex~$56\times 56$ matrices associated with its fundamental
representation is easily shown to be well beyond technological reach:
One would naively expect both the real and imaginary part of a typical
matrix element to be a sum of contributions containing a sine or
cosine factor in each coordinate, so a reasonable estimate for the
number of individual trigonometric factors needed to represent each
matrix element is~$2\cdot42\cdot2^{42}\approx4\cdot10^{14}$.  This
must be regarded as a very conservative lower estimate, considering
that that the occurrence of spinorial~$SO(8)$ representations the
definition of in~$E_7$ means that there will be more options than just
a~$\sin(\alpha)$ or $\cos(\alpha)$ factor for each angle coordinate
(the angle can occur with a number of different half-integral
coefficients): Clearly, a full analytic treatment of the problem hence
is technically unfeasible. What has been done so far to nevertheless
extract some information about the symmetry breaking structure of this
potential was to use a group-theoretic argument to reduce the analysis
to specially chosen more tractable low-dimensional sub-manifolds
of~$E_{7}/SU(8)$ that are invariant under some subgroup of the gauge
group~$SO(8)$. If this subgroup~$G$ is chosen sufficiently large, or
if a suitable embedding is chosen, the $G$-invariant
submanifold~$\mathcal{M}_G$ may allow a manageable analytic
parametrization.

While this strategy allowed the determination of a number of
nontrivial solutions, and by far has not been exploited to its full
theoretical potential yet, considering the large number of possible
options for~$G$ such that~$\mathcal{M}_G$ is low-dimensional and
easily parametrized, this approach seems to be somewhat unrewarding,
as for many specific choices of residual symmetry, the potential by no
means has to have associated stationary points. Still, it seems
reasonable to expect the number of stationary points with almost
completely -- or even completely -- broken gauge symmetry to be fairly
large. For this reason, other, complementary, techniques must be
considered to study the extremal structure of the scalar potentials of
models of supergravity -- this, as well as a number of others. As has
been demonstrated in~\cite{Fischbacher:2008zu}, the ``reverse mode
algorithmic differentiation'' method, also known as ``sensitivity
backpropagation'', is a stunningly potent numerical technique that
allows an extremely effective analysis of these problems: For the case
involving the largest exceptional Lie group~$E_8$, namely
$\mathcal{N}=16$ `Chern-Simons' supergravity in three dimensions with
gauge group $SO(8)\times SO(8)$, dozens of new solutions could be
determined with very little effort. It hence makes sense to also apply
this technique to gauged $\mathcal{N}=8$, $D=4$ supergravity.

\section{The scalars of $SO(8)$-gauged $\mathcal{N}=8$ SUGRA}

\noindent The deformation of maximal supergravity in four dimensions
that promotes the~$28$ vector fields to nonabelian gauge bosons
of~$SO(8)$ requires the introduction of a scalar potential at second
order in the coupling constant~$g$, in order to maintain
supersymmetry.  Due to the emergent global~$E_7$ symmetry of maximal
four-dimensional supergravity, this potential is most easily expressed
as a particular sixth-order polynomial function of the entries of
an~$E_7$ `$56$-bein' associated with the coset manifold of
scalars~$E_{7(7)}/SU(8)$. This function has a local maximum where all
scalar fields are set to zero, which is associated with
unbroken~$SO(8)$ gauge symmetry and full ~$\mathcal{N}=8$
supersymmetry in Anti-deSitter space with a negative cosmological
constant of~$V=-6g^2$. In addition to this, there are a number of
further saddle points in the potential, some of which nevertheless are
stable due to the strong gravitational back-reaction in such a
spacetime with Planck-scale cosmological constant: as long as the
energy gained by leaving a stationary point is over-compensated by the
energy that has to be expended in the kinetic term for a localized
variation, the solution is perturbatively stable. This stability
criterion is encoded in the Breitenlohner-Freedman 
bound~\cite{Breitenlohner:1982jf} on the lowest mass of the scalar
excitations:

\begin{equation}
\frac{m_{\rm min}^2/g^2}{-V/g^2}\ge-\frac{9}{4}=-2.25.
\end{equation}

When listing a number of stationary points with highly broken gauge
symmetry, such as here, it makes sense to spell out the conventions
on~$E_7$ matrices in full detail and then refer to this common basis,
rather than discussing each case in terms of the types of scalar
fields that get excited and their interrelation. This is done in the
rest of this section.

\subsection{The $E_{7(+7)}$ Lie algebra}

\noindent The Lie algebra of the group~$E_{7(7)}$ allows a beautiful
triality-symmetric construction based on the subalgebra~$so(8)$: Just
as $so(8)$ can be extended to $so(9)$ by adding eight extra generators
that transform in the eight-dimensional vector representation ${\bf
  8}_{\rm v}$, plus the obvious commutators for these new generators
amongst one another, we can also perform a triality-symmetric
extension of $so(8)$ to $f_4$ by simultaneously adding the three
inequivalent real eight-dimensional vector~(V), spinor~(S), and
co-spinor~(C) representations, again with the obvious commutator
relations, which here are e.g. of the form $[V,S]=C$ and involve
the~$so(8)$ Clifford algebra generators~$\gamma^\va_{\sb\cc}$. If one
instead suitably extends~$so(8)$ by the symmetric traceless matrices
over the vectors, spinors, and co-spinors, ${\bf 35}_{\rm v}$, ${\bf
  35}_{\rm s}$, ${\bf 35}_{\rm c}$, one obtains $e_7$. Signs and
factors of~$i$ can be chosen in such a way that the generators
obtained from ${\bf 28}+{\bf 35}_{\rm v}$ give rise to the Lie algebra
of the maximal compact subgroup,~$su(8)$, while ${\bf 35}_{\rm s}+{\bf
  35}_{\rm c}$ give the 70 `boost' generators of $E_{7(7)}$.  In terms
of the maximal compact~$SU(8)$ subgroup of~$E_{7(7)}$, the ${\bf
  35}_{\rm s}$ and ${\bf 35}_{\rm c}$ correspond to the self-dual and
anti-self-dual four-forms.

A similar triality-symmetric construction of $e_{8(8)}$ starts from
the subalgebra~$so(8)\times so(8)$ and extends this with three
64-dimensional representations that transform as $({\bf 8}_{\rm v},
{\bf 8}_{\rm v})$, $({\bf 8}_{\rm s}, {\bf 8}_{\rm s})$, and $({\bf
  8}_{\rm c}, {\bf 8}_{\rm c})$. The maximal split form is obtained by
choosing signs in such a way that $so(8)\times so(8)+({\bf 8}_{\rm v},
{\bf 8}_{\rm v})$ form the Lie algebra of the maximal compact
subgroup,~$so(16)$.

It is useful to note that, when taking a diagonal~$so(8)$ subalgebra
of~$so(8)\times so(8)$, these 64-dimensional representations each
decompose as~${\bf 28}_{\rm anti}+({\bf 1}+{\bf 35})_{\rm symm}$, and
the diagonal~$so(8)$ together with the three~${\bf 35}$ give
$e_{7(7)}$, while the three singlets form the~$sl(2)$ of the maximal
subalgebra~$e_{7(7)}\times sl(2)$ of $e_{8(8)}$.

From this discussion, we see that explicit generators of~$E_7$ that
are constructed along these lines are slightly more awkward to work
with than the corresponding generators of~$E_8$, the reason being that
the most obvious and convenient choices of a (rational) basis for the
${\bf 35}_{\rm v,s,c}$ representations are non-orthonormal. This also
has direct implications for strategies to simplify the presentation of
stationary points that have been found numerically.

In the following, we will use different alphabets to designate
different irreducible representations. Note that for some
non-fundamental representations, we use composite symbols which
nevertheless are supposed to be read as single glyphs. The notation
may at times seem unusually tedious, but was chosen with the objective
in mind to simplify transliteration of formulas to computer code as
much as possible -- with the exception of \emph{not} adopting the
convention that index counting starts at zero, which, while highly
desirable from a technical perspective, would be too much in
disagreement with established practice in physics. The alphabets
listed in table~\ref{tab:indices} will be employed.  We also
occasionally use primes to enlarge alphabets, hence both $\va$ and
$\va'$ will be considered to be $so(8)$ vector indices. Furthermore,
composite indices such as $\underline{[ijk]}$ are used for
lexicographically enumerated 3-forms.

\begin{table}
\begin{center}
\begin{tabular}{|l|l|l|}
\hline
Representation&Indices&Range\\
\hline
\hline
$so(8)$ vector&$\va,\;\vb,\;\ldots$&$1\ldots 8$\\
$so(8)$ spinor&$\sa,\;\sb,\;\ldots$&$1\ldots 8$\\
$so(8)$ co-spinor&$\ca,\;\cb,\;\ldots$&$1\ldots 8$\\
$so(8)\;{\bf 35}_s$&$\sstab,\;\sstcd\;\ldots$&$1\ldots 35$\\
$so(8)\;{\bf 35}_c$&$\sctab,\;\sctcd\;\ldots$&$1\ldots 35$\\
$su(8)$ vector&$\ua,\;\ub,\;\ldots;\uA,\;\uB,\;\ldots$&$1\ldots 8$\\
$su(8)$ adjoint&$\Ua,\;\Ub,\;\ldots$&$1\ldots 63$\\
$e_7$ fundamental&$\efa,\;\efb,\;\ldots$&$1\ldots 56$\\
$e_7$ adjoint&$\eaa,\;\eab,\;\ldots$&$1\ldots 133$\\
\hline
None (counting index)&$\ma,\;\mb,\;\ldots$&Situation-dependent\\
\hline
\end{tabular}
\caption{\label{tab:indices}Index alphabets}
\end{center}
\end{table}

The ``typewriter indices'' $\ma,\;\mb,\;\ldots$ that occasionally show
up serve a dual purpose: on the one hand, they allow us to take some
short-cuts, such as when defining index range decompositions.  In
particular, they are employed e.g. to avoid the otherwise common
`matrix block notation' for decomposing the complex $56\times 56$
matrices that represent $E_{7(7)}$ group elements, which may be
dangerous here due to possible factor-2 ambiguities. Their second
purpose is to keep expressions computer friendly and allow a more or
less straightforward transliteration of these formulas to computer
code.\footnote{If index counting uniformly starts at zero, 
rather than one, this makes conversion of tensor multi-indices to
linear indices much more straightforward. Hence, it generally is
advisable to adhere to this convention in computer calculations, 
and shift indices when producing \LaTeX{} output.}

Also, we will frequently make use of the `lecicographical index
splitting function'~$Z(\mn)$ that maps a number~$\mn$ in the
range~$1\ldots 28$ to an index pair~$(\mi,\mj)$, $1\le \mi<\mj\le 8$ 
such that~$1$ gets mapped to~$(1,2)$, $2$~to~$(1,3)$, $7$~to~$(1,8)$,
$8$~to~$(2,3)$, etc.: $Z((i-1)\cdot8+j-i(i+1)/2)=(i,j)$.

The detailed construction of~$133$ complex $56\times 56$ generator
matrices that satisfy the~$e_{7}$ commutator algebra is given
in~\ref{sec:appendix-e7}.

\subsection{The scalar potential}

\noindent Parametrizing the coset manifold of
scalars~$\mathcal{H}=E_{7}/SU(8)$ by~$70$ boost generator
coefficients~$\phi_\mn$, the potential~$V(\phi_\mn)$ of~$SO(8)$-gauged
$\mathcal{N}=8, D=4$~supergravity is a simple quadratic expression in
the tensors~$A_1$, $A_2$, which are defined in terms of the so-called
$T$-tensor that is (in~$D=4$) cubic in entries of the exponentiated
generator~$\exp\left(\sum_\mn\phi_\mn g^{(\mn)}\right)$, specifically:
%

\begin{equation}
\label{eq:potential}
\begin{array}{lcl}
V/g^2&=&-\frac{3}{4}A_1^{ij}\left(A_1^{ij}\right)^*+\frac{1}{24}A_2{}^{i}{}_{jkl}\left(A_2{}^{i}{}_{jkl}\right)^*\\
\mbox{with:}&&\\
A_1{}^{ij}&=&-\frac{4}{21}T_m{}^{ijm}\\
A_{2\,\ell}{}^{ijk}&=&-\frac{4}{3}T_{\ell}{}^{i'j'k'}\delta_{i'j'k'}^{ijk}\\
T_{\ell}{}^{kij}&=&\left(u^{ij}{}_{IJ}+v^{ijIJ}\right)\left(u_{\ell m}{}^{JK}u^{km}{}_{KI}-v_{\ell mJK}v^{kmKI}\right)\\
&&\\
\mathcal{V}^\efa{}_\efb&=&\exp\left(\sum_\mn\phi_\mn g^{(\mn)}\right)^\efa{}_\efb\\
u_{ij}{}^{IJ}&=&2\,\mathcal{V}^\efa{}_\efb\,\delta_\efa^{\mm}\delta^\efb_{\mn}\delta_{ij}^{\ma\mb}\delta^{IJ}_{\mc\md}\\
&\mbox{for}&\efa\le 28,\;\efb\le 28,\;(\ma,\mb)=Z(\mm),\;(\mc,\md)=Z(\mn)\\
u^{kl}{}_{KL}&=&2\,\mathcal{V}^\efa{}_\efb\,\delta_\efa^{\mm}\delta^\efb_{\mn}\delta^{kl}_{\ma\mb}\delta_{KL}^{\mc\md}\\
&\mbox{for}&\efa>28,\;\efb>28,\;(\ma,\mb)=Z(\mm-28),\;(\mc,\md)=Z(\mn-28)\\
v_{ijKL}&=&2\,\mathcal{V}^\efa{}_\efb\,\delta_\efa^{\mm}\delta^\efb_{\mn}\delta_{ij}^{\ma\mb}\delta_{KL}^{\mc\md}\\
&\mbox{for}&\efa\le28,\;\efb>28,\;(\ma,\mb)=Z(\mm),\;(\mc,\md)=Z(\mn-28)\\

v^{klIJ}&=&2\,\mathcal{V}^\efa{}_\efb\,\delta_\efa^{\mm}\delta^\efb_{\mn}\delta^{kl}_{\ma\mb}\delta^{IJ}_{\mc\md}\\
&\mbox{for}&\efa>28,\;\efb\le28,\;(\ma,\mb)=Z(\mm-28),\;(\mc,\md)=Z(\mn)
\end{array}
\end{equation}

In this work, all anti-symmetrizers are normalized to work as
projectors, e.g.~$\delta^{abc}_{def}=\delta^{abc}_{ghi}\delta^{ghi}_{def}$
and likewise for the projectors $\delta^{ab}_{cd}$ and
$\delta^{abcd}_{efgh}$:
\begin{equation}
\delta^{a_1a_2\ldots a_n}_{b_1b_2\ldots b_n}=\left\{\begin{array}{lcl}
+1/n!&\mbox{for}& b_1b_2\ldots b_n\;\;\mbox{an even permutation of}\;\;a_1a_2\ldots a_n,\\
-1/n!&\mbox{for}& b_1b_2\ldots b_n\;\;\mbox{an odd permutation of}\;\;a_1a_2\ldots a_n,\\
0&\mbox{else}&\\
\end{array}\right.
\end{equation}

\section{Previously known stationary points}

\noindent The potential~$V(\phi_\mn)$ defined in~(\ref{eq:potential})
is known to have a local maximum at~$\phi_\mn=0$ with~$V=-6g^2$ of
maximal symmetry~$SO(8)$ and unbroken~$\mathcal{N}=8$ supersymmetry.
In addition to this, six further stationary points in this potential
have been published, which are also shown in
table~\ref{tab:KnownSolutions}.  Their detailed locations are given
in~\cite{Warner:1983vz,Warner:1983du}.  Numerical data are given
in~\ref{sec:appendix-locations} in particular to simplify matching the
conventions used in this work against conventions used in other parts
of the literature.

\begin{table}
\begin{center}
{\small
\begin{tabular}{|lllcc|}
\hline
Nr.&$V(\phi_\mn)/g^2$&approx.&Residual gauge group&Residual SUSY\\
\hline
0&-\phantom06\phantom{.000000}&{\tt -\phantom06\phantom{.000000}}&$SO(8)$&$\mathcal{N}=8$\\
\hline
1&$-2\cdot5^{3/4}$&{\tt -\phantom06.687403}&$SO(7)_{s}$&--\\
2&$-5^{5/2}\cdot2^{-3}$&{\tt -\phantom06.987712}&$SO(7)_{c}$&--\\
3&$-2^{7/2}\cdot3^{13/4}\cdot5^{-5/2}$&{\tt -\phantom07.191576}&$G_{2}$&$\mathcal{N}=1$\\
4&$-3^{5/2}\cdot 2^{-1}$&{\tt -\phantom07.794229}&$SU(3)\times U(1)$&$\mathcal{N}=2$\\
5&$-\phantom08\phantom{.000000}$&{\tt -\phantom08\phantom{.000000}}&$SU(4)$&--\\
6&$-14\phantom{.000000}$&{\tt -14\phantom{.000000}}&$SO(3)\times SO(3)$&--\\
\hline
\end{tabular}
}
\caption{\label{tab:KnownSolutions}The previously known stationary points}
\end{center}
\end{table}

\section{New stationary points}

\noindent Employing the sensitivity backpropagation method presented
in~\cite{Fischbacher:2008zu}, which already demonstrated its utility
for the study of the somewhat more involved~$E_8$ potential
of~$\mathcal{N}=16$ three-dimensional Chern-Simons supergravity with
gauge group~$SO(8)\times SO(8)$, strong numerical evidence for a
number of further stationary points in the potential could be
obtained. Here, it is important to point out that, while more than
doubling the previously known amount of data, the new stationary
points listed below perhaps are just a comparatively small selection
of the totality of further solutions. While the present article
demonstrates the utility of this numerical technique, a more complete
analysis will be deferred to a subsequent publication that also
addresses a number of open issues with respect to more convenient and
hence usable presentations of solutions, as well as semi-automatic
derivation of robust analytical expressions (guided by numerical
input).  Essentially, it is feasible to use heuristics based on
numerical observations to set up algebraic equation systems that both
can be solved exactly and also stringently checked against the
stationarity conditions. This is briefly explained in the last section
of~\cite{FischbacherValidation}. The new stationary points are given
in table~\ref{tab:NewSolutions}.

\begin{table}
\begin{center}
\begin{tabular}{|llcc|}
\hline
Nr.&approx. $V(\phi_\mn)/g^2$&Residual gauge group&Residual SUSY\\
\hline
\hline
\phantom07&{\tt -\phantom09.987083}&$U(1)$&--\\
\phantom08&{\tt -10.434713}&--&--\\
\phantom09&{\tt -10.674754}&$U(1)\times U(1)$&--\\
10&{\tt -11.656854}&$U(1)\times U(1)$&--\\
11&{\tt -12.000000}&$U(1)\times U(1)$&$\mathcal{N}=1$?\\
12&{\tt -13.623653}&$U(1)$&--\\
13&{\tt -13.676114}&--&--\\
14&{\tt -14.970385}&$U(1)$&--\\
15&{\tt -16.414456}&--&--\\
16&{\tt -17.876428}&--&--\\
17&{\tt -18.052693}&--&--\\
18&{\tt -21.265976}&--&--\\
19&{\tt -21.408498}&--&--\\
20&{\tt -25.149369}&--&--\\
\hline
\end{tabular}
\caption{\label{tab:NewSolutions}The new stationary points}
\end{center}
\end{table}

Detailed information on their location as well as their residual
symmetries and fermion masses are given in~\ref{sec:appendix-locations}.

\section{Details on the calculation}

\noindent For these investigations, the ``misalignment function'' on
the scalar manifold to be minimized numerically has been taken to be
the length-squared of the potential's gradient, rather than an
explicit quadratic stationarity condition in the $A$-tensors, as
in~\cite{Fischbacher:2008zu}. The corresponding $A_1/A_2$ stationarity
condition is listed nevertheless~(cf.  e.g.~$(2.21)$ and~$(2.22)$, as
well as the associated discussion in~\cite{deWit:1983gs}) and has been
used as an independent check to numerically validate the new solutions:
\begin{equation}
\begin{array}{lcl}
Q^{ijkl}&=&-\frac{1}{24}\epsilon^{ijklmnpq} Q_{mnpq}\\
\mbox{with}&&\\
Q^{ijkl}&=&\left(\frac{3}{4}A_{2\,m}{}^{ni'j'}A_{2\,n}{}^{k'l'm}-A_1{}^{mi'}A_{2\,m}{}^{j'k'l'}\right)\delta^{ijkl}_{i'j'k'l'}
\end{array}
\end{equation}

This decision was motivated in part by the desire to learn whether the
computation of quantities involving second derivatives would be
feasible using the sensitivity backpropagation method, and how good
its performance would be. As it turns out, the computations for this
potential on average take considerably longer than the corresponding
calculations for the~$E_8$ potential of three-dimensional maximal
gauged supergravity, but still are quite feasible even on a single
notebook computer. Also, it is noteworthy that, with this approach,
the time needed to obtain a solution as well as the numerical accuracy
that is easily obtainable vary much stronger than in the
three-dimensional case studied in~\cite{Fischbacher:2008zu}. This
might in part be attributed to the potential being a sixth-order
polynomial in the entries of the $T$-tensor here, while it is only
fourth order in~$D=3$. While the superiority of this numerical method
for identifying stationary points is again clearly demonstrated, a
subsequent more detailed analysis that intends to determine a large
number of solutions should perhaps use code that is better optimized
and uses the $Q$-tensor criterion. For this reason, no code is yet
included in the {\tt arxiv.org} preprint upload of this work.

A technical issue arises concerning the presentation of results:
typically, a solution found numerically will initially be given in a
fairly awkward way, a more or less generic numerical vector of
length~70. In many cases, in particular in situations with large
residual gauge symmetries, this may be simplified considerably by
choosing a suitable coordinate basis (i.e. finding an
appropriate~$SO(8)$ rotation) that sets many of the entries to zero.
Due to the `diagonal traceless' parts of the~${\bf 35}_s$ and~${\bf
  35}_c$ representations, this is most easily implemented in the
language of four-forms, obtained by\footnote{This fully real
  parametrization differs from the usual (complex) four-form
  parametrization, but for the objective of finding a rotation that
  sets many entries to zero, this does not make much of a
  difference. This particular choice makes the sensitivity 
  backpropagation code a bit easier to implement.}:
\begin{equation}\label{eq:phi4}
\phi_{\va\vb\vc\vd}=
\sum_{\mn=1}^{35} S^{(SO(8))}{}_{\sstab}^{\va\vb\vc\vd}\,\phi_\mn\delta^\mn_\sstab
+\sum_{\mn=36}^{70} C^{(SO(8))}{}_{\sctab}^{\va\vb\vc\vd}\,\phi_\mn\delta^{\mn-35}_\sctab
\end{equation}

The ``niceness''$N(\phi)$ of a presentation of a~$70$-vector is then obtained
by taking:
\begin{equation}\label{eq:phi4opt}
N(\phi)=\sum_{\va\vb\vc\vd}\left(\phi_{\va\vb\vc\vd}\right)^4.
\end{equation}

Using the sensitivity backpropagation method again, a~$SO(8)$ rotation
that maximizes this niceness can be found quite efficiently. This
`niceness' function has no physical significance and is inspired by
the simple observation that, when rotating a 2-dimensional vector
$(v_1,v_2)$ whose direction can be chosen freely, setting one
coordinate to zero is equivalent to maximizing $v_1^4+v_2^4$ (any even
positive exponent larger than 2 would also do).

It is somewhat puzzling, however, to observe that, while this strategy
clearly manages to simplify results, it also seems to often fail to
come up with the best possible presentation here. In particular, when
doing the numerical analysis, it seemed to produce one fairly awkward
presentation for the $SU(4)$-symmetric stationary point as it was
re-discovered using these methods. Hence, a more exhaustive analysis
should also pay more attention to this issue.

\section{Discussion and Outlook}

\noindent While having only numerical data on the locations of these
new stationary points clearly is somewhat unsatisfying, it is equally
clear that these numerical techniques provide highly valuable clues
that should make it possible to analytically determine the exact
properties of many of the new solutions. Taking as a specific example
the numerical data on the location of the stationary point
with~$V/g^2=-13.676114$, \emph{which does not seem to have any
  residual gauge symmetry}, an ansatz such this one seems highly
suggestive:

\begin{equation}
\mbox{\begin{minipage}{10cm}
$-\phi_{[1257]+}=+\phi_{[1457]+}=-\phi_{[2368]+}=+\phi_{[3468]+}=A$,\\
$-\phi_{[1348]+}=+\phi_{[1478]+}=+\phi_{[2356]+}=-\phi_{[2567]+}=B$,\\
$+\phi_{[1568]+}=+\phi_{[2347]+}=C$,\\
$+\phi_{[1235]+}=-\phi_{[1345]+}=+\phi_{[2678]+}=-\phi_{[4678]+}=D$,\\
$-\phi_{[1246]+}=+\phi_{[1367]+}=-\phi_{[2458]+}=+\phi_{[3578]+}=E$,\\
$+\phi_{[1235]-}=-\phi_{[1345]-}=-\phi_{[2678]-}=+\phi_{[4678]-}=F$,\\
$-\phi_{[1348]-}=-\phi_{[1478]-}=+\phi_{[2356]-}=+\phi_{[2567]-}=G$,\\
$+\phi_{[1257]-}=-\phi_{[1457]-}=-\phi_{[2368]-}=+\phi_{[3468]-}=H$,\\
$-\phi_{[1246]-}=+\phi_{[1367]-}=+\phi_{[2458]-}=-\phi_{[3578]-}=I$,\\
$-\phi_{[1238]-}=-\phi_{[1278]-}=+\phi_{[3456]-}=+\phi_{[4567]-}=J$,\\
$-\phi_{[1238]+}=+\phi_{[1278]+}=+\phi_{[3456]+}=-\phi_{[4567]+}=K$,\\

$A\approx0.0210$, $B\approx0.0403$, $C\approx0.0863$, $D\approx0.0961$,\\
 $E\approx0.1024$, $F\approx0.1221$, $G\approx0.1488$, $H\approx0.1970$,\\
 $I\approx0.3271$, $J\approx0.6484$, $K\approx0.6513$.
\end{minipage}}
\end{equation}

This alone considerably reduces the complexity of the problem. Here,
one has to keep in mind that the large number of stationary points to
be investigated strongly favours (semi-)automizable strategies over
strategies that involve manual computations.  A promising route to
continue from here is to next determine the parameters~$A-K$ to high
accuracy, using either fast high-precision floatingpoint arithmetics
as provided e.g. by~\cite{Hida}, or more conventional implementations
of multiprecision arithmetics. This then should suffice to derive
additional hypotheses on algebraic relations between these parameters.
While this should further reduce the number of variables and hence
simplify analytic matrix exponentiation, we note in passing that full
analytic (`Euler angle') parametrizations of nine-dimensional
submanifolds of~$E_7$ have already been technically feasible seven
years ago. Likewise, hybrid symbolic-numeric techniques can be
employed to automatically generate and verify conjectures for
algebraic relations between the entries of~$A_1$ and~$A_2$, and hence
ultimately the matrix~$\mathcal{V}^{\efa}{}_{\efb}$.

Therefore, a complete determination of the exact properties of a
substantial fraction of all the stationary points of the scalar
potential of~$\mathcal{N}=8$, $D=4$, $SO(8)$-gauged supergravity
should have come within technological reach now.

Considering the specific data that have been produced in this
investigation, the most exciting outcome is of course the numerical
evidence for a new vacuum with residual~$U(1)\times U(1)$ gauge
symmetry that seems to have a residual~$\mathcal{N}=1$ supersymmetry.
Using numerical data for the $A_1$ and $A_2$ tensors that can be
obtained with the code provided in~\cite{FischbacherValidation}, this
can be seen in two ways: on the one hand, the $A_1$ tensor is almost
diagonal, with just a $2\times 2$ block in the (7,8)-coordinates that
is of the form:
\begin{equation}
M=\left(\begin{array}{lcl}
2.1213251&0.7071019\\
0.7071019&2.1213251
\end{array}\right)
\end{equation}

Hence, the vector $\epsilon$ with $\epsilon_j=\delta_j^7-\delta_j^8$
gets mapped to $1.4142232\epsilon$; the square of this eigenvalue is
$2.0000273$. According to \cite{Gibbons:1983aq} (see also 
(3.13)--(3.15) in \cite{Nicolai:1985hs}), any eigenvector of $A_1$ 
for which the corresponding eigenvalue $\lambda$ satisfies 
$\lambda^2=-V/6g^2$ (here: $\lambda^2=2$) corresponds to a
residual supersymmetry.  Also, the approximate numerical data given 
for $A_2$ seem to support $\epsilon_i A_2{}^i{}_{jkl}=0$, hence 
providing independent numerical evidence.

The numerical data available so far already suggests a fairly simple
structure that might be the basis of an analytic investigation:

\begin{equation}
\begin{array}{l}
\phantom=-\phi_{[1357]+}=-\phi_{[1468]+}=+\phi_{[2357]+}=-\phi_{[2468]+}=\\
=-\phi_{[1357]-}=-\phi_{[1468]-}=-\phi_{[2357]-}=+\phi_{[2468]-}=A,\\
\\
\phantom=+\phi_{[1367]+}=-\phi_{[1458]+}=-\phi_{[2367]+}=-\phi_{[2458]+}=\\
=-\phi_{[1367]-}=+\phi_{[1458]-}=-\phi_{[2367]-}=-\phi_{[2458]-}=B,\\
\\
-\phi_{[1278]-}=+\phi_{[3456]-}=C\\
\\
A\approx0.0013, B\approx0.5731, C\approx0.6585
\end{array}
\end{equation}

Further work on this solution will show whether this simple
three-parameter ansatz can be established analytically.

\appendix
\renewcommand{\theequation}{\Alph{section}.\arabic{equation}}
\renewcommand{\thesection}{Appendix \Alph{section}}

\section{$E_{7(7)}$ conventions}\label{sec:appendix-e7}

We start the triality-inspired construction of explicit~$E_{7(7)}$
generators by explicitly stating the entries of the~$so(8)$ 
invariants $\gamma^\va{}_{\sa\ca}$. The choice given here
reproduces the conventions in \cite{Green:1987sp}: the nonzero 
entries of $\gamma^\va{}_{\sa\ca}$ are listed in compact
notation as $PQR_\pm$, meaning
$\gamma^{\va=P}{}_{\sa=Q\,\ca=R}=\pm1$:

\begin{equation}
\begin{array}{llllllll}
118_{+}&127_{-}&136_{-}&145_{+}&154_{-}&163_{+}&172_{+}&181_{-}\\
212_{+}&221_{-}&234_{-}&243_{+}&256_{+}&265_{-}&278_{-}&287_{+}\\
315_{+}&326_{-}&337_{+}&348_{-}&351_{-}&362_{+}&373_{-}&384_{+}\\
413_{+}&424_{+}&431_{-}&442_{-}&457_{-}&468_{-}&475_{+}&486_{+}\\
514_{+}&523_{-}&532_{+}&541_{-}&558_{+}&567_{-}&576_{+}&585_{-}\\
616_{+}&625_{+}&638_{+}&647_{+}&652_{-}&661_{-}&674_{-}&683_{-}\\
717_{+}&728_{+}&735_{-}&746_{-}&753_{+}&764_{+}&771_{-}&782_{-}\\
811_{+}&822_{+}&833_{+}&844_{+}&855_{+}&866_{+}&877_{+}&888_{+}
\end{array}
\end{equation}

From these, we then construct:
\begin{eqnarray}
\gamma^{\va\vb\vc\vd}_{\alpha\beta}&=&\gamma^\ve_{\alpha\dot\gamma}\gamma^\vf_{\gamma\dot\gamma}\gamma^\vg_{\gamma\dot\epsilon}\gamma^\vh_{\beta\dot\epsilon}\delta^{\va\vb\vc\vd}_{\ve\vf\vg\vh}\\
\gamma^{\va\vb\vc\vd}_{\dot\alpha\dot\beta}&=&\gamma^\ve_{\gamma\dot\alpha}\gamma^\vf_{\gamma\dot\gamma}\gamma^\vg_{\epsilon\dot\gamma}\gamma^\vh_{\epsilon\dot\beta}\delta^{\va\vb\vc\vd}_{\ve\vf\vg\vh}
\end{eqnarray}

With these conventions, the tensor entries of~$\gamma^{\va\vb\vc\vd}_{\alpha\beta}$
and~$\gamma^{\va\vb\vc\vd}_{\dot\alpha\dot\beta}$ are $\pm 1$~and~$0$.

The choice of basis for the 133-dimensional $E_{7(7)}$ Lie-algebra is
such that the last 28 basis elements (i.e. elements 105--133) spawn
the subalgebra of $SO(8)$. Elements~1--35 transform under this~$SO(8)$
as symmetric traceless (s.t.) matrices over the spinors, elements~36--70 as
s.t.~matrices over the co-spinors, and elements~71--105
as s.t.~matrices over the vectors. Hence, elements~71--133 form the
subalgebra of~$SU(8)$, and elements~1--70 are the `boost' generators
associated with the~$70$ scalars of supergravity that live on the coset
manifold~$E_{7(7)}/SU(8)$. Hence, an~$E_7$ adjoint index~$\mathcal A$
splits as:
\begin{equation}
\mathcal{A}\rightarrow \underline{(\alpha\beta)} + \underline{(\dot\alpha\dot\beta)} + \underline{(ij)} + \underline{[ij]}
\end{equation}

We furthermore choose the basis for~$SO(8)$ in such a way that
element~$105+n$, when acting on the vector representation of~$SO(8)$,
would be represented as the rotation
matrix~$(R_{\underline{[jk]}})^m{}_n=\delta^j_m \delta^k_n-\delta^j_n
\delta^k_m$, i.e. the rotation that takes the~$k$-direction into
the~$j$-direction, with~$1\le j\le k\le 8$,
$n=(j-1)\cdot8+(k-1)-(j-1)j/2$. Hence,~$E_7$ basis 
element~$\#106$ corresponds to the rotation~$R_{\underline{[12]}}$,
basis element~$107=R_{\underline{[13]}}$, etc. (lexicographically).

For the three $35$-dimensional symmetric traceless
vector/spinor/co-spinor representations, we use the convention that
the first~$7$ basis elements correspond to the diagonal 
matrices~$\rm diag(1,-1,0,\ldots,0)$, $\rm diag(0,1,-1,0,\ldots,0)$,
$\rm diag(0,\ldots,0,1,-1)$ (in that order), while element~$7+n$
corresponds to the matrix~$(S_{\underline{(jk)}})^m{}_p=\delta^j_m \delta^k_p+\delta^j_p\delta^k_m$, again with~$1\le j\le k\le 8$,
$n=(j-1)\cdot8+(k-1)-(j-1)j/2$ -- and with a likewise
lexicographical order for the corresponding non-diagonal
parts of~$S_{\underline{(\alpha\beta)}}$
and~$S_{\underline{(\dot\alpha\dot\beta)}}$.
These conventions ensure that the~$E_7$
symmetric bilinear form obtained from the fundamental
representation,~$G_{\eaa\eab}=T_{\eaa\efc}{}^\efd T_{\eab\efd}{}^\efc$
is almost diagonal, with entries~$+96$ for~$\efb=\efc\le70$,
entries~$-96$ for~$\efb=\efc>70$, entries~$-48$ for
the non-orthogonal generators corresponding to the diagonal
parts of the symmetric traceless matrices over the spinors
and co-spinors
(i.e. $G_{\eaa=1\,\eab=2},\;G_{\eaa=4\,\eab=3}\;G_{\eaa=36\,\eab=37}$,
etc.), and entries~$+48$ for $G_{\eaa=71\,\eab=72}$, etc., which
come from the diagonal part of the~${\bf 35}_v$ representation.

In order to obtain explicit complex~$56\times 56$ matrices for~$E_7$
generators in the fundamental representation, we first define the
tensors below (where the Einstein summation convention is suspended for
`technical' auxiliary indices that are set in typewriter font and do
not belong to irreducible representations). Note that the adjoint
representation of~$SU(8)$ in the ``bottom right'' $28\times 28$ matrix
block involves complex conjugation, at variance with some of the
earlier literature on this subject. Without this, the 133-dimensional
algebra would not close. Evidently, a convenient choice for a Cartan
subalgebra of this~$E_7$ algebra is given by the
generators~$\#71,\ldots,\#77$. Given~$T^{(E7)}{}_{\eaa\efb}{}^{\efc}$,
the~$E_7$ generator matrices~$g$ used to define the scalar potential
are~$\left(g^{(\mn)}\right)^{\efc}{}_{\efb}=\left(T^{(E7)}{}_{\eaa=\mn}\right)_{\efb}{}^{\efc}$
for~$n=1,\ldots,70$. 

With our conventions, the structure of the $56\times56$ generator
matrices $\left(g^{(\mn)}\right)$ is as follows: For $\mn\le 35$, each
matrix has either zero or two non-zero entries per row which then are
either $+1$ or $-1$, the total number of non-zero rows being 48.  For
$35<\mn\le 70$, each matrix likewise has 48 rows with two non-zero
entries each that are either $+i$ or $-i$. For $70<\mn\le105$, each
matrix has 24 non-zero entries in total, which each are either $+2i$
or $-2i$.  Finally, for $\mn>105$, each matrix has 24 non-zero
entries, with each of them being either $+2$ or $-2$. In total, there
are $70\cdot48\cdot2+63\cdot24=8232$ non-zero entries in the
tensor~$T^{(E7)}$.

\begin{equation}
\begin{array}{lcl}
T^{(SU(8))}{}_{\Ua\ub}{}^\uc&=&\left\{
\begin{array}{lcl}
+i&\mbox{for}&\Ua=\ub=\uc\le 7\\
-i&\mbox{for}&\Ua+1=\ub=\uc\le 7\\
+i&\mbox{for}&7<\Ua\le35,\;(\mm,\mn)=Z(\Ua-7), \ub=\mm, \uc=\mn\\
+i&\mbox{for}&7<\Ua\le35,\;(\mm,\mn)=Z(\Ua-7), \ub=\mn, \uc=\mm\\
+1&\mbox{for}&35<\Ua,\; (\mm,\mn)=Z(\Ua-35), \ub=\mm, \uc=\mn\\
-1&\mbox{for}&35<\Ua,\; (\mm,\mn)=Z(\Ua-35), \ub=\mn, \uc=\mm\\
\end{array}\right.\\
&&\\
S^{(SO(8))}{}_{\sstab}^{\va\vb\vc\vd}&=&\left\{
\begin{array}{lcl}
\gamma^{\va\vb\vc\vd}_{\sa\sb}(
\delta^\sa_{\mm}\delta^\sb_{\mm}
-\delta^\sa_{\mn}\delta^\sb_{\mn})
&\mbox{for}&\sstab=\mm=\mn-1\le7\\
\gamma^{\va\vb\vc\vd}_{\sa\sb}(\delta^\sa_{\mm}\delta^\sb_{\mn}+\delta^\sa_{\mn}\delta^\sb_{\mm})&\mbox{for}&\sstab>7,\;{(\mm,\mn)}=Z(\sstab-7)
\end{array}\right.\\
&&\\
C^{(SO(8))}{}_\sctab^{\va\vb\vc\vd}&=&\left\{
\begin{array}{lcl}
\gamma^{\va\vb\vc\vd}_{\ca\cb}(
\delta^{\ca}_{\mm}\delta^{\cb}_{\mm}
-\delta^{\ca}_{\mn}\delta^{\cb}_{\mn})
&\mbox{for}&\sctab=\mm=\mn-1\le7\\
\gamma^{\va\vb\vc\vd}_{\ca\cb}(\delta^{\ca}_{\mm}\delta^{\cb}_{\mn}+\delta^{\ca}_{\mn}\delta^{\cb}_{\mm})&\mbox{for}&\sctab>7,\;{(\mm,\mn)}=Z(\sctab-7)
\end{array}\right.\\
&&\\
T^{(E7)}{}_{\eaa\efb}{}^{\efc}&=&\left\{
\begin{array}{lcl}
\frac{1}{8}S^{(SO(8))}{}_{\sstab}^{\va\vb\vc\vd}(\delta^\va_\mm\delta^\vb_\mn-\delta^\va_\mn\delta^\vb_\mm)(\delta^\vc_\mp\delta^\vd_\mq-\delta^\vc_\mq\delta^\vd_\mp)\qquad\mbox{for}&&\\
\qquad\eaa\le35,\;\sstab=\eaa,&&\\
\qquad\efb>28,\;(\mp,\mq)=Z(\efb-28),&&\\
\qquad\efc\le28,\;(\mm,\mn)=Z(\efc)&&\\
\hline\\
\frac{1}{8}S^{(SO(8))}{}_{\sstab}^{\vc\vd\va\vb}(\delta^\vc_\mm\delta^\vd_\mn-\delta^\vc_\mn\delta^\vd_\mm)(\delta^\va_\mp\delta^\vb_\mq-\delta^\va_\mq\delta^\vb_\mp)\qquad\mbox{for}&&\\
\qquad\eaa\le35,\;\sstab=\eaa,&&\\
\qquad\efb\le28,\;(\mp,\mq)=Z(\efb),&&\\
\qquad\efc>28,\;(\mm,\mn)=Z(\efc-28)&&\\
\hline\\
\frac{i}{8}C^{(SO(8))}{}_{\sctab}^{\va\vb\vc\vd}(\delta^\va_\mm\delta^\vb_\mn-\delta^\va_\mn\delta^\vb_\mm)(\delta^\vc_\mp\delta^\vd_\mq-\delta^\vc_\mq\delta^\vd_\mp)\qquad\mbox{for}&&\\
\qquad35<\eaa\le70,\;\sctab=\eaa-35,&&\\
\qquad\efb>28,\;(\mp,\mq)=Z(\efb-28),&&\\
\qquad\efc\le28,\;(\mm,\mn)=Z(\efc)&&\\
\hline\\
-\frac{i}{8}C^{(SO(8))}{}_{\sctab}^{\vc\vd\va\vb}(\delta^\vc_\mm\delta^\vd_\mn-\delta^\vc_\mn\delta^\vd_\mm)(\delta^\va_\mp\delta^\vb_\mq-\delta^\va_\mq\delta^\vb_\mp)\qquad\mbox{for}&&\\
\qquad35<\eaa\le70,\;\sctab=\eaa-35,&&\\
\qquad\efb\le28,\;(\mp,\mq)=Z(\efb),&&\\
\qquad\efc>28,\;(\mm,\mn)=Z(\efc-28)&&\\
\hline\\
2\,T^{(SU(8))}{}_{\Ua\ub}{}^\uc\left(%
  \delta^\mq_\mn\delta^\ub_\mm\delta_\uc^\mp
 +\delta^\mp_\mm\delta^\ub_\mn\delta_\uc^\mq
 -\delta^\mp_\mn\delta^\ub_\mm\delta_\uc^\mq
 -\delta^\mq_\mm\delta^\ub_\mn\delta_\uc^\mp
\right) \qquad\mbox{for}&&\\
\qquad\eaa>70,\;\Ua=\eaa-70&&\\
\qquad\efb\le28,\;(\mp,\mq)=Z(\efb),&&\\
\qquad\efc\le28,\;(\mm,\mn)=Z(\efc)&&\\
\hline\\
2\,\left(T^{(SU(8))}{}_{\Ua\ub}{}^\uc\right)^*\left(%
  \delta^\mq_\mn\delta^\ub_\mm\delta_\uc^\mp
 +\delta^\mp_\mm\delta^\ub_\mn\delta_\uc^\mq
 -\delta^\mp_\mn\delta^\ub_\mm\delta_\uc^\mq
 -\delta^\mq_\mm\delta^\ub_\mn\delta_\uc^\mp
\right) \qquad\mbox{for}&&\\
\qquad\eaa>70,\;\Ua=\eaa-70&&\\
\qquad\efb>28,\;(\mp,\mq)=Z(\efb-28),&&\\
\qquad\efc>28,\;(\mm,\mn)=Z(\efc-28)&&\\
\end{array}\right.\end{array}
\end{equation}

\section{Numerical data}\label{sec:appendix-locations}

\noindent This section lists approximate numerical data both for the
known as well as for the new solution. Each table gives:
\begin{itemize}
  \item The identifier of the solution (matching
    tables~\ref{tab:KnownSolutions} and~\ref{tab:NewSolutions})
  \item The value of the scalar potential $V/g^2$
  \item The `quality' of the solution, $|Q|$ being the Frobenius
    norm of the $Q$-tensor and $|\nabla|$ being the length of the
    gradient (determined numerically via finite
    differences)\footnote{The validation code presented in
      \cite{FischbacherValidation} with which these tables were
      produced does not contain sensitivity backpropagation;
      accurate gradients that were determined via backpropagation
      show that the stationarity condition is satisfied even better
      than the numbers given here indicate. Floatingpoint accuracy
      sets a limit of $|\nabla|\ge10^{-5.9}$ here, as can be seen from
      solution {\bf \#0}.}
  \item The residual symmetry and supersymmetry.
  \item The approximate location of the (degenerate) solution.
    For convenience, the data obtained using equations~\ref{eq:phi4}
    and~\ref{eq:phi4opt} have been post-processed into a more usual 
    scalar(+)/pseudoscalar(-) coefficient notation.
  \item Approximate data on residual gauge group generators, for
    solutions with $U(1)^n$ symmetry.

  \item The (re-scaled) masses-squared of the fermions. 
    For a (potentially complex) symmetric fermion mass matrix
    $M$, these are given by the eigenvalues of $M^* M$.
    All eigenvalues have been re-scaled by a factor 
    $1/m_0^2=-6/(V/g^2)$.

    The gravitino masses-squared $(m^2/m_0^2)[\psi]$ are
    the eigenvalues of $A_1^* A_1$, while the spin-$1/2$ fermion
    masses-squared are the  eigenvalues of $A_3^* A_3$, with~$A_3$
    being the matrix:
\begin{equation}
\begin{array}{lcl}
  A_3{}_{\underline{[ijk]}\,\underline{[lmn]}}&=&
\frac{\sqrt{2}}{144}\epsilon_{ijkpqrlm}A_2{}_n{}^{pqr}P_{\underline{[ijk]}}^{ijk}P_{\underline{[lmn]}}^{lmn}\\
&=&-\frac{\sqrt{2}}{108}\epsilon_{ijkpqrlm}T_n{}^{pqr}P_{\underline{[ijk]}}^{ijk}P_{\underline{[lmn]}}^{lmn}
\end{array}
\end{equation}
    (Cf. eq.~(2.14) in~\cite{deWit:1983gs} and eq.~(5.23) 
    in~\cite{de Wit:1982ig}).
    Here, $P_{\underline{[ijk]}}^{ijk}$ is the 
    $56\times 8\times 8\times 8$ tensor that is fully 
    anti-symmetric in $ijk$ and has entries 
    $\pm1, 0$. Entries $\pm1$ occur for $ijk$ being a 
    cyclic/anti-cyclic permutation of the $\underline{[ijk]}$-th 
    index triplet without repetition in lexicographical order.
\end{itemize}

A companion article~\cite{FischbacherValidation} provides numerical
computer code containing more accurate data for the locations of all
stationary points in the extended list, code to validate claims about
stationarity, about residual (super-)symmetry, code to re-produce
these tables, and also code to study the properties of these
stationary points in more detail.

\begin{longtable}{||l||}
\hline
\raise-0.1ex\hbox{\small {\bf \#0}: $V/g^2=-6.000000$, Quality: $|Q|=10^{-\infty}$, $|\nabla|=10^{-5.86}$  $\mathcal{N}=8$}\\
\hline
$\phi$\quad\begin{minipage}{10.5cm}
{\vskip0.25ex
\small\tt
0
}\end{minipage}\\
\hline
Symmetry\quad\begin{minipage}{8.5cm}
{\vskip0.25ex
\small\tt
[28-dimensional]
}\end{minipage}\\
\hline

$(m^2/m_0^2)[\psi]$\quad\begin{minipage}{8.5cm}
{\vskip0.25ex
\small
{\tt 1.000}$_{(\times 8)}$
}\end{minipage}\\
\hline
$(m^2/m_0^2)[\chi]$\quad\begin{minipage}{8.5cm}
{\vskip0.25ex
\small
{\tt 0.000}$_{(\times 56)}$
}\end{minipage}\\
\hline
\end{longtable}

\begin{longtable}{||l||}
\hline
\raise-0.1ex\hbox{\small {\bf \#1}: $V/g^2=-6.687403$, Quality: $|Q|=10^{-14.67}$, $|\nabla|=10^{-5.04}$ }\\
\hline
$\phi$\quad\begin{minipage}{10.5cm}
{\vskip0.25ex
\small\tt
$+0.2012_{[1234]+}$, $-0.2012_{[1256]+}$, $-0.2012_{[1278]+}$, $-0.2012_{[1358]+}$, $-0.2012_{[1367]+}$, $+0.2012_{[1457]+}$, $-0.2012_{[1468]+}$, $+0.2012_{[2357]+}$, $-0.2012_{[2368]+}$, $+0.2012_{[2458]+}$, $+0.2012_{[2467]+}$, $-0.2012_{[3456]+}$, $-0.2012_{[3478]+}$, $+0.2012_{[5678]+}$
}\end{minipage}\\
\hline
Symmetry\quad\begin{minipage}{8.5cm}
{\vskip0.25ex
\small\tt
[21-dimensional]
}\end{minipage}\\
\hline

$(m^2/m_0^2)[\psi]$\quad\begin{minipage}{8.5cm}
{\vskip0.25ex
\small
{\tt 1.350}$_{(\times 8)}$
}\end{minipage}\\
\hline
$(m^2/m_0^2)[\chi]$\quad\begin{minipage}{8.5cm}
{\vskip0.25ex
\small
{\tt 2.700}$_{(\times 8)}$, {\tt 0.075}$_{(\times 48)}$
}\end{minipage}\\
\hline
\end{longtable}

\begin{longtable}{||l||}
\hline
\raise-0.1ex\hbox{\small {\bf \#2}: $V/g^2=-6.987712$, Quality: $|Q|=10^{-14.30}$, $|\nabla|=10^{-5.06}$ }\\
\hline
$\phi$\quad\begin{minipage}{10.5cm}
{\vskip0.25ex
\small\tt
$+0.2406_{[1234]-}$, $-0.2406_{[1256]-}$, $+0.2406_{[1278]-}$, $+0.2406_{[1358]-}$, $-0.2406_{[1367]-}$, $+0.2406_{[1457]-}$, $+0.2406_{[1468]-}$, $+0.2406_{[2357]-}$, $+0.2406_{[2368]-}$, $-0.2406_{[2458]-}$, $+0.2406_{[2467]-}$, $-0.2406_{[3456]-}$, $+0.2406_{[3478]-}$, $-0.2406_{[5678]-}$
}\end{minipage}\\
\hline
Symmetry\quad\begin{minipage}{8.5cm}
{\vskip0.25ex
\small\tt
[21-dimensional]
}\end{minipage}\\
\hline

$(m^2/m_0^2)[\psi]$\quad\begin{minipage}{8.5cm}
{\vskip0.25ex
\small
{\tt 1.350}$_{(\times 8)}$
}\end{minipage}\\
\hline
$(m^2/m_0^2)[\chi]$\quad\begin{minipage}{8.5cm}
{\vskip0.25ex
\small
{\tt 2.700}$_{(\times 8)}$, {\tt 0.075}$_{(\times 48)}$
}\end{minipage}\\
\hline
\end{longtable}

\begin{longtable}{||l||}
\hline
\raise-0.1ex\hbox{\small {\bf \#3}: $V/g^2=-7.191576$, Quality: $|Q|=10^{-5.18}$, $|\nabla|=10^{-4.56}$  $\mathcal{N}=1$}\\
\hline
$\phi$\quad\begin{minipage}{10.5cm}
{\vskip0.25ex
\small\tt
$+0.1457_{[1234]+}$, $-0.1457_{[1256]+}$, $-0.1457_{[1278]+}$, $-0.1457_{[1358]+}$, $-0.1457_{[1367]+}$, $+0.1457_{[1457]+}$, $-0.1457_{[1468]+}$, $+0.1457_{[2357]+}$, $-0.1457_{[2368]+}$, $+0.1457_{[2458]+}$, $+0.1457_{[2467]+}$, $-0.1457_{[3456]+}$, $-0.1457_{[3478]+}$, $+0.1457_{[5678]+}$, $+0.2139_{[1234]-}$, $-0.2139_{[1256]-}$, $+0.2139_{[1278]-}$, $+0.2139_{[1358]-}$, $-0.2139_{[1367]-}$, $+0.2139_{[1457]-}$, $+0.2139_{[1468]-}$, $+0.2139_{[2357]-}$, $+0.2139_{[2368]-}$, $-0.2139_{[2458]-}$, $+0.2139_{[2467]-}$, $-0.2139_{[3456]-}$, $+0.2139_{[3478]-}$, $-0.2139_{[5678]-}$
}\end{minipage}\\
\hline
Symmetry\quad\begin{minipage}{8.5cm}
{\vskip0.25ex
\small\tt
[14-dimensional]
}\end{minipage}\\
\hline

$(m^2/m_0^2)[\psi]$\quad\begin{minipage}{8.5cm}
{\vskip0.25ex
\small
{\tt 1.500}$_{(\times 7)}$, {\tt 1.000}
}\end{minipage}\\
\hline
$(m^2/m_0^2)[\chi]$\quad\begin{minipage}{8.5cm}
{\vskip0.25ex
\small
{\tt 3.000}$_{(\times 8)}$, {\tt 0.750}$_{(\times 7)}$, {\tt 0.083}$_{(\times 27)}$, {\tt 0.000}$_{(\times 14)}$
}\end{minipage}\\
\hline
\end{longtable}

\begin{longtable}{||l||}
\hline
\raise-0.1ex\hbox{\small {\bf \#4}: $V/g^2=-7.794229$, Quality: $|Q|=10^{-14.40}$, $|\nabla|=10^{-4.39}$  $\mathcal{N}=2$}\\
\hline
$\phi$\quad\begin{minipage}{10.5cm}
{\vskip0.25ex
\small\tt
$+0.2747_{[1234]+}$, $+0.2747_{[1256]+}$, $+0.2747_{[1278]+}$, $+0.2747_{[3456]+}$, $+0.2747_{[3478]+}$, $+0.2747_{[5678]+}$, $+0.3292_{[1357]-}$, $-0.3292_{[1368]-}$, $-0.3292_{[1458]-}$, $-0.3292_{[1467]-}$, $+0.3292_{[2358]-}$, $+0.3292_{[2367]-}$, $+0.3292_{[2457]-}$, $-0.3292_{[2468]-}$
}\end{minipage}\\
\hline
Symmetry\quad\begin{minipage}{8.5cm}
{\vskip0.25ex
\small\tt
[9-dimensional]
}\end{minipage}\\
\hline

$(m^2/m_0^2)[\psi]$\quad\begin{minipage}{8.5cm}
{\vskip0.25ex
\small
{\tt 1.778}$_{(\times 6)}$, {\tt 1.000}$_{(\times 2)}$
}\end{minipage}\\
\hline
$(m^2/m_0^2)[\chi]$\quad\begin{minipage}{8.5cm}
{\vskip0.25ex
\small
{\tt 3.556}$_{(\times 6)}$, {\tt 3.281}$_{(\times 2)}$, {\tt 1.219}$_{(\times 2)}$, {\tt 0.889}$_{(\times 12)}$, {\tt 0.056}$_{(\times 18)}$, {\tt 0.000}$_{(\times 16)}$
}\end{minipage}\\
\hline
\end{longtable}

\begin{longtable}{||l||}
\hline
\raise-0.1ex\hbox{\small {\bf \#5}: $V/g^2=-8.000000$, Quality: $|Q|=10^{-14.52}$, $|\nabla|=10^{-4.49}$ }\\
\hline
$\phi$\quad\begin{minipage}{10.5cm}
{\vskip0.25ex
\small\tt
$+0.4407_{[1357]-}$, $-0.4407_{[1368]-}$, $-0.4407_{[1458]-}$, $-0.4407_{[1467]-}$, $+0.4407_{[2358]-}$, $+0.4407_{[2367]-}$, $+0.4407_{[2457]-}$, $-0.4407_{[2468]-}$
}\end{minipage}\\
\hline
Symmetry\quad\begin{minipage}{8.5cm}
{\vskip0.25ex
\small\tt
[15-dimensional]
}\end{minipage}\\
\hline

$(m^2/m_0^2)[\psi]$\quad\begin{minipage}{8.5cm}
{\vskip0.25ex
\small
{\tt 1.688}$_{(\times 8)}$
}\end{minipage}\\
\hline
$(m^2/m_0^2)[\chi]$\quad\begin{minipage}{8.5cm}
{\vskip0.25ex
\small
{\tt 3.375}$_{(\times 8)}$, {\tt 2.344}$_{(\times 8)}$, {\tt 0.094}$_{(\times 40)}$
}\end{minipage}\\
\hline
\end{longtable}

\begin{longtable}{||l||}
\hline
\raise-0.1ex\hbox{\small {\bf \#6}: $V/g^2=-14.000000$, Quality: $|Q|=10^{-14.58}$, $|\nabla|=10^{-4.27}$ }\\
\hline
$\phi$\quad\begin{minipage}{10.5cm}
{\vskip0.25ex
\small\tt
$+1.0208_{[1235]+}$, $-1.0208_{[4678]+}$, $+1.0208_{[1234]-}$, $-1.0208_{[5678]-}$
}\end{minipage}\\
\hline
Symmetry\quad\begin{minipage}{8.5cm}
{\vskip0.25ex
\small\tt
[6-dimensional]
}\end{minipage}\\
\hline

$(m^2/m_0^2)[\psi]$\quad\begin{minipage}{8.5cm}
{\vskip0.25ex
\small
{\tt 3.857}$_{(\times 2)}$, {\tt 2.143}$_{(\times 6)}$
}\end{minipage}\\
\hline
$(m^2/m_0^2)[\chi]$\quad\begin{minipage}{8.5cm}
{\vskip0.25ex
\small
{\tt 7.714}$_{(\times 2)}$, {\tt 4.286}$_{(\times 12)}$, {\tt 2.244}$_{(\times 18)}$, {\tt 0.327}$_{(\times 18)}$, {\tt 0.000}$_{(\times 6)}$
}\end{minipage}\\
\hline
\end{longtable}

\begin{longtable}{||l||}
\hline
\raise-0.1ex\hbox{\small {\bf \#7}: $V/g^2=-9.987083$, Quality: $|Q|=10^{-5.77}$, $|\nabla|=10^{-4.26}$ $U(1)$}\\
\hline
$\phi$\quad\begin{minipage}{10.5cm}
{\vskip0.25ex
\small\tt
$-0.1493_{[1235]+}$, $-0.0094_{[1238]+}$, $+0.1792_{[1246]+}$, $+0.0237_{[1257]+}$, $+0.5627_{[1278]+}$, $-0.0648_{[1345]+}$, $-0.1578_{[1348]+}$, $+0.1719_{[1367]+}$, $+0.2089_{[1457]+}$, $-0.0651_{[1478]+}$, $-0.1792_{[1568]+}$, $-0.1792_{[2347]+}$, $-0.0651_{[2356]+}$, $-0.2089_{[2368]+}$, $-0.1719_{[2458]+}$, $-0.1578_{[2567]+}$, $+0.0648_{[2678]+}$, $+0.5627_{[3456]+}$, $-0.0237_{[3468]+}$, $-0.1792_{[3578]+}$, $-0.0094_{[4567]+}$, $+0.1493_{[4678]+}$, $+0.4970_{[1235]-}$, $-0.0599_{[1238]-}$, $+0.1113_{[1246]-}$, $-0.0772_{[1257]-}$, $+0.0316_{[1278]-}$, $-0.1270_{[1345]-}$, $+0.4271_{[1348]-}$, $-0.5403_{[1367]-}$, $+0.0358_{[1457]-}$, $+0.1338_{[1478]-}$, $+0.1113_{[1568]-}$, $-0.1113_{[2347]-}$, $-0.1338_{[2356]-}$, $+0.0358_{[2368]-}$, $-0.5403_{[2458]-}$, $-0.4271_{[2567]-}$, $-0.1270_{[2678]-}$, $-0.0316_{[3456]-}$, $-0.0772_{[3468]-}$, $+0.1113_{[3578]-}$, $+0.0599_{[4567]-}$, $+0.4970_{[4678]-}$
}\end{minipage}\\
\hline
Symmetry\quad\begin{minipage}{8.5cm}
{\vskip0.25ex
\small\tt
$+0.399\,R_{13}$ $-0.088\,R_{17}$ $-0.361\,R_{25}$ $-0.435\,R_{28}$ $+0.088\,R_{36}$ $+0.376\,R_{45}$ $+0.454\,R_{48}$ $-0.399\,R_{67}$
}\end{minipage}\\
\hline

$(m^2/m_0^2)[\psi]$\quad\begin{minipage}{8.5cm}
{\vskip0.25ex
\small
{\tt 2.797}$_{(\times 2)}$, {\tt 2.197}$_{(\times 2)}$, {\tt 2.081}, {\tt 1.890}$_{(\times 2)}$, {\tt 1.598}
}\end{minipage}\\
\hline
$(m^2/m_0^2)[\chi]$\quad\begin{minipage}{8.5cm}
{\vskip0.25ex
\small
{\tt 5.594}$_{(\times 2)}$, {\tt 4.394}$_{(\times 2)}$, {\tt 4.162}, {\tt 4.116}$_{(\times 2)}$, {\tt 3.861}$_{(\times 2)}$, {\tt 3.781}$_{(\times 2)}$, {\tt 3.770}, {\tt 3.745}, {\tt 3.196}, {\tt 2.564}$_{(\times 2)}$, {\tt 2.301}, {\tt 1.896}$_{(\times 2)}$, {\tt 1.454}$_{(\times 2)}$, {\tt 1.400}$_{(\times 2)}$, {\tt 1.381}$_{(\times 2)}$, {\tt 1.375}, {\tt 0.795}$_{(\times 2)}$, {\tt 0.584}, {\tt 0.508}$_{(\times 2)}$, {\tt 0.225}, {\tt 0.201}$_{(\times 2)}$, {\tt 0.155}, {\tt 0.153}, {\tt 0.131}$_{(\times 2)}$, {\tt 0.124}$_{(\times 2)}$, {\tt 0.104}$_{(\times 2)}$, {\tt 0.098}$_{(\times 2)}$, {\tt 0.080}$_{(\times 2)}$, {\tt 0.077}, {\tt 0.049}$_{(\times 2)}$, {\tt 0.040}$_{(\times 2)}$, {\tt 0.038}$_{(\times 2)}$, {\tt 0.029}$_{(\times 2)}$, {\tt 0.017}
}\end{minipage}\\
\hline
\end{longtable}

\begin{longtable}{||l||}
\hline
\raise-0.1ex\hbox{\small {\bf \#8}: $V/g^2=-10.434713$, Quality: $|Q|=10^{-7.38}$, $|\nabla|=10^{-4.41}$ }\\
\hline
$\phi$\quad\begin{minipage}{10.5cm}
{\vskip0.25ex
\small\tt
$+0.2101_{[1235]+}$, $-0.2101_{[1238]+}$, $+0.1924_{[1246]+}$, $+0.4126_{[1257]+}$, $+0.1426_{[1278]+}$, $+0.0677_{[1345]+}$, $+0.0677_{[1348]+}$, $+0.2293_{[1367]+}$, $+0.4126_{[1457]+}$, $-0.1426_{[1478]+}$, $-0.1924_{[1568]+}$, $-0.1924_{[2347]+}$, $-0.1426_{[2356]+}$, $-0.4126_{[2368]+}$, $-0.2293_{[2458]+}$, $+0.0677_{[2567]+}$, $-0.0677_{[2678]+}$, $+0.1426_{[3456]+}$, $-0.4126_{[3468]+}$, $-0.1924_{[3578]+}$, $-0.2101_{[4567]+}$, $-0.2101_{[4678]+}$, $+0.2807_{[1235]-}$, $-0.2807_{[1238]-}$, $-0.3141_{[1246]-}$, $-0.4490_{[1257]-}$, $+0.0339_{[1278]-}$, $-0.1600_{[1345]-}$, $-0.1600_{[1348]-}$, $+0.4490_{[1457]-}$, $+0.0339_{[1478]-}$, $+0.3141_{[1568]-}$, $-0.3141_{[2347]-}$, $-0.0339_{[2356]-}$, $+0.4490_{[2368]-}$, $+0.1600_{[2567]-}$, $-0.1600_{[2678]-}$, $-0.0339_{[3456]-}$, $-0.4490_{[3468]-}$, $-0.3141_{[3578]-}$, $+0.2807_{[4567]-}$, $+0.2807_{[4678]-}$
}\end{minipage}\\
\hline
$(m^2/m_0^2)[\psi]$\quad\begin{minipage}{8.5cm}
{\vskip0.25ex
\small
{\tt 3.023}, {\tt 2.620}$_{(\times 2)}$, {\tt 2.330}, {\tt 2.241}, {\tt 1.951}$_{(\times 2)}$, {\tt 1.651}
}\end{minipage}\\
\hline
$(m^2/m_0^2)[\chi]$\quad\begin{minipage}{8.5cm}
{\vskip0.25ex
\small
{\tt 6.047}, {\tt 5.240}$_{(\times 2)}$, {\tt 4.661}, {\tt 4.483}, {\tt 4.056}, {\tt 4.005}, {\tt 3.940}$_{(\times 2)}$, {\tt 3.901}$_{(\times 2)}$, {\tt 3.782}, {\tt 3.628}$_{(\times 2)}$, {\tt 3.302}, {\tt 3.035}, {\tt 3.014}, {\tt 2.576}$_{(\times 2)}$, {\tt 2.073}, {\tt 1.644}, {\tt 1.498}$_{(\times 2)}$, {\tt 1.395}, {\tt 1.142}, {\tt 1.052}$_{(\times 2)}$, {\tt 0.946}, {\tt 0.896}, {\tt 0.789}$_{(\times 2)}$, {\tt 0.648}, {\tt 0.613}, {\tt 0.463}$_{(\times 2)}$, {\tt 0.411}, {\tt 0.140}, {\tt 0.118}$_{(\times 2)}$, {\tt 0.107}, {\tt 0.106}, {\tt 0.081}, {\tt 0.078}, {\tt 0.071}, {\tt 0.057}$_{(\times 2)}$, {\tt 0.049}$_{(\times 2)}$, {\tt 0.032}$_{(\times 2)}$, {\tt 0.031}, {\tt 0.022}, {\tt 0.006}, {\tt 0.002}$_{(\times 2)}$, {\tt 0.001}
}\end{minipage}\\
\hline
\end{longtable}

\begin{longtable}{||l||}
\hline
\raise-0.1ex\hbox{\small {\bf \#9}: $V/g^2=-10.674754$, Quality: $|Q|=10^{-4.12}$, $|\nabla|=10^{-3.77}$ $U(1)\times U(1)$}\\
\hline
$\phi$\quad\begin{minipage}{10.5cm}
{\vskip0.25ex
\small\tt
$+0.1682_{[1235]+}$, $+0.2182_{[1246]+}$, $-0.4136_{[1248]+}$, $-0.0518_{[1267]+}$, $-0.1797_{[1278]+}$, $-0.0517_{[1346]+}$, $-0.1801_{[1348]+}$, $+0.2176_{[1367]+}$, $-0.4137_{[1378]+}$, $-0.1958_{[1457]+}$, $-0.2316_{[1568]+}$, $-0.2316_{[2347]+}$, $+0.1958_{[2368]+}$, $+0.4137_{[2456]+}$, $-0.2176_{[2458]+}$, $-0.1801_{[2567]+}$, $-0.0517_{[2578]+}$, $-0.1797_{[3456]+}$, $-0.0518_{[3458]+}$, $+0.4136_{[3567]+}$, $-0.2182_{[3578]+}$, $-0.1682_{[4678]+}$, $+0.3291_{[1246]-}$, $+0.3290_{[1248]-}$, $-0.0003_{[1267]-}$, $-0.0003_{[1346]-}$, $-0.3287_{[1367]-}$, $-0.3288_{[1378]-}$, $+0.6578_{[1457]-}$, $+0.6578_{[2368]-}$, $-0.3288_{[2456]-}$, $-0.3287_{[2458]-}$, $+0.0003_{[2578]-}$, $+0.0003_{[3458]-}$, $+0.3290_{[3567]-}$, $+0.3291_{[3578]-}$
}\end{minipage}\\
\hline
Symmetry\quad\begin{minipage}{8.5cm}
{\vskip0.25ex
\small\tt
$-0.500\,R_{12}$ $+0.500\,R_{25}$ $-0.500\,R_{67}$ $+0.500\,R_{78}$; $-0.354\,R_{14}$ $-0.354\,R_{17}$ $+0.353\,R_{26}$ $+0.354\,R_{28}$ $-0.354\,R_{36}$ $+0.353\,R_{38}$ $-0.354\,R_{45}$ $-0.354\,R_{57}$
}\end{minipage}\\
\hline

$(m^2/m_0^2)[\psi]$\quad\begin{minipage}{8.5cm}
{\vskip0.25ex
\small
{\tt 2.656}$_{(\times 4)}$, {\tt 2.137}$_{(\times 4)}$
}\end{minipage}\\
\hline
$(m^2/m_0^2)[\chi]$\quad\begin{minipage}{8.5cm}
{\vskip0.25ex
\small
{\tt 5.312}$_{(\times 4)}$, {\tt 4.273}$_{(\times 4)}$, {\tt 3.995}$_{(\times 4)}$, {\tt 3.673}$_{(\times 4)}$, {\tt 3.463}$_{(\times 4)}$, {\tt 1.344}$_{(\times 4)}$, {\tt 1.195}$_{(\times 4)}$, {\tt 0.802}$_{(\times 4)}$, {\tt 0.638}$_{(\times 4)}$, {\tt 0.177}$_{(\times 4)}$, {\tt 0.090}$_{(\times 4)}$, {\tt 0.086}$_{(\times 4)}$, {\tt 0.085}$_{(\times 4)}$, {\tt 0.001}$_{(\times 4)}$
}\end{minipage}\\
\hline
\end{longtable}

\begin{longtable}{||l||}
\hline
\raise-0.1ex\hbox{\small {\bf \#10}: $V/g^2=-11.656854$, Quality: $|Q|=10^{-6.50}$, $|\nabla|=10^{-4.13}$ $U(1)\times U(1)$}\\
\hline
$\phi$\quad\begin{minipage}{10.5cm}
{\vskip0.25ex
\small\tt
$-0.3330_{[1238]+}$, $+0.3330_{[1245]+}$, $+0.3330_{[1356]+}$, $+0.3751_{[1368]+}$, $+0.3330_{[1468]+}$, $-0.3330_{[2357]+}$, $+0.3751_{[2457]+}$, $-0.3330_{[2478]+}$, $+0.3330_{[3678]+}$, $-0.3330_{[4567]+}$, $-0.1841_{[1235]-}$, $-0.3268_{[1238]-}$, $-0.3268_{[1245]-}$, $-0.5802_{[1248]-}$, $+0.3268_{[1356]-}$, $-0.5802_{[1368]-}$, $-0.1841_{[1456]-}$, $+0.3268_{[1468]-}$, $+0.3268_{[2357]-}$, $+0.1841_{[2378]-}$, $+0.5802_{[2457]-}$, $+0.3268_{[2478]-}$, $-0.5802_{[3567]-}$, $+0.3268_{[3678]-}$, $+0.3268_{[4567]-}$, $-0.1841_{[4678]-}$
}\end{minipage}\\
\hline
Symmetry\quad\begin{minipage}{8.5cm}
{\vskip0.25ex
\small\tt
$-0.574\,R_{18}$ $-0.413\,R_{24}$ $+0.413\,R_{36}$ $-0.574\,R_{57}$; $-0.413\,R_{18}$ $+0.574\,R_{24}$ $-0.574\,R_{36}$ $-0.413\,R_{57}$
}\end{minipage}\\
\hline

$(m^2/m_0^2)[\psi]$\quad\begin{minipage}{8.5cm}
{\vskip0.25ex
\small
{\tt 2.561}$_{(\times 8)}$
}\end{minipage}\\
\hline
$(m^2/m_0^2)[\chi]$\quad\begin{minipage}{8.5cm}
{\vskip0.25ex
\small
{\tt 5.121}$_{(\times 8)}$, {\tt 3.685}$_{(\times 8)}$, {\tt 3.464}$_{(\times 8)}$, {\tt 0.854}$_{(\times 8)}$, {\tt 0.802}$_{(\times 8)}$, {\tt 0.118}$_{(\times 8)}$, {\tt 0.002}$_{(\times 8)}$
}\end{minipage}\\
\hline
\end{longtable}

\begin{longtable}{||l||}
\hline
\raise-0.1ex\hbox{\small {\bf \#11}: $V/g^2=-12.0$, Quality: $|Q|=10^{-6.09}$, $|\nabla|=10^{-4.29}$ $U(1)\times U(1)$ $\mathcal{N}=1$}\\
\hline
$\phi$\quad\begin{minipage}{10.5cm}
{\vskip0.25ex
\small\tt
$-0.0013_{[1357]+}$, $+0.5731_{[1367]+}$, $-0.5731_{[1458]+}$, $-0.0013_{[1468]+}$, $+0.0013_{[2357]+}$, $-0.5731_{[2367]+}$, $-0.5731_{[2458]+}$, $-0.0013_{[2468]+}$, $-0.6585_{[1278]-}$, $-0.0013_{[1357]-}$, $-0.5731_{[1367]-}$, $+0.5731_{[1458]-}$, $-0.0013_{[1468]-}$, $-0.0013_{[2357]-}$, $-0.5731_{[2367]-}$, $-0.5731_{[2458]-}$, $+0.0013_{[2468]-}$, $+0.6585_{[3456]-}$
}\end{minipage}\\
\hline
Symmetry\quad\begin{minipage}{8.5cm}
{\vskip0.25ex
\small\tt
$+0.707\,R_{36}$ $-0.707\,R_{45}$; $-0.707\,R_{36}$ $-0.707\,R_{45}$
}\end{minipage}\\
\hline

$(m^2/m_0^2)[\psi]$\quad\begin{minipage}{8.5cm}
{\vskip0.25ex
\small
{\tt 4.000}, {\tt 3.000}$_{(\times 2)}$, {\tt 2.250}$_{(\times 4)}$, {\tt 1.000}
}\end{minipage}\\
\hline
$(m^2/m_0^2)[\chi]$\quad\begin{minipage}{8.5cm}
{\vskip0.25ex
\small
{\tt 8.000}, {\tt 6.000}$_{(\times 2)}$, {\tt 4.747}$_{(\times 4)}$, {\tt 4.500}$_{(\times 4)}$, {\tt 3.937}$_{(\times 4)}$, {\tt 3.732}, {\tt 2.166}$_{(\times 4)}$, {\tt 2.000}$_{(\times 5)}$, {\tt 1.500}$_{(\times 2)}$, {\tt 1.125}$_{(\times 4)}$, {\tt 0.584}$_{(\times 4)}$, {\tt 0.500}$_{(\times 4)}$, {\tt 0.268}, {\tt 0.125}$_{(\times 4)}$, {\tt 0.064}$_{(\times 4)}$, {\tt 0.003}$_{(\times 4)}$, {\tt 0.000}$_{(\times 4)}$
}\end{minipage}\\
\hline
\end{longtable}

\begin{longtable}{||l||}
\hline
\raise-0.1ex\hbox{\small {\bf \#12}: $V/g^2=-13.623653$, Quality: $|Q|=10^{-5.90}$, $|\nabla|=10^{-4.43}$ $U(1)$}\\
\hline
$\phi$\quad\begin{minipage}{10.5cm}
{\vskip0.25ex
\small\tt
$+0.3511_{[1234]+}$, $+0.5011_{[1256]+}$, $-0.1538_{[1278]+}$, $+0.0188_{[1358]+}$, $+0.4933_{[1367]+}$, $-0.4933_{[1457]+}$, $+0.0188_{[1468]+}$, $-0.0188_{[2357]+}$, $+0.4933_{[2368]+}$, $-0.4933_{[2458]+}$, $-0.0188_{[2467]+}$, $-0.1538_{[3456]+}$, $+0.5011_{[3478]+}$, $+0.3511_{[5678]+}$, $+0.1309_{[1234]-}$, $+0.5531_{[1256]-}$, $+0.3862_{[1278]-}$, $-0.1728_{[1358]-}$, $+0.5789_{[1367]-}$, $-0.5789_{[1457]-}$, $-0.1728_{[1468]-}$, $-0.1728_{[2357]-}$, $-0.5789_{[2368]-}$, $+0.5789_{[2458]-}$, $-0.1728_{[2467]-}$, $-0.3862_{[3456]-}$, $-0.5531_{[3478]-}$, $-0.1309_{[5678]-}$
}\end{minipage}\\
\hline
Symmetry\quad\begin{minipage}{8.5cm}
{\vskip0.25ex
\small\tt
$+0.707\,R_{34}$ $+0.707\,R_{56}$
}\end{minipage}\\
\hline

$(m^2/m_0^2)[\psi]$\quad\begin{minipage}{8.5cm}
{\vskip0.25ex
\small
{\tt 3.425}, {\tt 3.391}, {\tt 3.377}, {\tt 3.081}$_{(\times 2)}$, {\tt 2.307}$_{(\times 2)}$, {\tt 1.431}
}\end{minipage}\\
\hline
$(m^2/m_0^2)[\chi]$\quad\begin{minipage}{8.5cm}
{\vskip0.25ex
\small
{\tt 6.850}, {\tt 6.783}, {\tt 6.755}, {\tt 6.162}$_{(\times 2)}$, {\tt 5.621}, {\tt 5.388}, {\tt 4.865}$_{(\times 2)}$, {\tt 4.862}$_{(\times 2)}$, {\tt 4.614}$_{(\times 2)}$, {\tt 4.367}, {\tt 4.271}, {\tt 4.046}, {\tt 3.403}$_{(\times 2)}$, {\tt 2.862}, {\tt 2.756}, {\tt 2.754}$_{(\times 2)}$, {\tt 2.284}, {\tt 2.004}$_{(\times 2)}$, {\tt 1.937}, {\tt 1.901}$_{(\times 2)}$, {\tt 1.553}, {\tt 1.505}$_{(\times 2)}$, {\tt 1.350}, {\tt 1.346}$_{(\times 2)}$, {\tt 1.272}, {\tt 0.957}$_{(\times 2)}$, {\tt 0.540}$_{(\times 2)}$, {\tt 0.305}, {\tt 0.242}$_{(\times 2)}$, {\tt 0.172}$_{(\times 2)}$, {\tt 0.088}$_{(\times 2)}$, {\tt 0.070}, {\tt 0.067}$_{(\times 2)}$, {\tt 0.043}$_{(\times 2)}$, {\tt 0.028}$_{(\times 2)}$, {\tt 0.019}, {\tt 0.008}, {\tt 0.005}
}\end{minipage}\\
\hline
\end{longtable}

\begin{longtable}{||l||}
\hline
\raise-0.1ex\hbox{\small {\bf \#13}: $V/g^2=-13.676114$, Quality: $|Q|=10^{-6.24}$, $|\nabla|=10^{-3.98}$ }\\
\hline
$\phi$\quad\begin{minipage}{10.5cm}
{\vskip0.25ex
\small\tt
$+0.0961_{[1235]+}$, $-0.6513_{[1238]+}$, $-0.1024_{[1246]+}$, $-0.0210_{[1257]+}$, $+0.6513_{[1278]+}$, $-0.0961_{[1345]+}$, $-0.0403_{[1348]+}$, $+0.1024_{[1367]+}$, $+0.0210_{[1457]+}$, $+0.0403_{[1478]+}$, $+0.0863_{[1568]+}$, $+0.0863_{[2347]+}$, $+0.0403_{[2356]+}$, $-0.0210_{[2368]+}$, $-0.1024_{[2458]+}$, $-0.0403_{[2567]+}$, $+0.0961_{[2678]+}$, $+0.6513_{[3456]+}$, $+0.0210_{[3468]+}$, $+0.1024_{[3578]+}$, $-0.6513_{[4567]+}$, $-0.0961_{[4678]+}$, $+0.1221_{[1235]-}$, $-0.6484_{[1238]-}$, $-0.3271_{[1246]-}$, $+0.1970_{[1257]-}$, $-0.6484_{[1278]-}$, $-0.1221_{[1345]-}$, $-0.1488_{[1348]-}$, $+0.3271_{[1367]-}$, $-0.1970_{[1457]-}$, $-0.1488_{[1478]-}$, $+0.1488_{[2356]-}$, $-0.1970_{[2368]-}$, $+0.3271_{[2458]-}$, $+0.1488_{[2567]-}$, $-0.1221_{[2678]-}$, $+0.6484_{[3456]-}$, $+0.1970_{[3468]-}$, $-0.3271_{[3578]-}$, $+0.6484_{[4567]-}$, $+0.1221_{[4678]-}$
}\end{minipage}\\
\hline
$(m^2/m_0^2)[\psi]$\quad\begin{minipage}{8.5cm}
{\vskip0.25ex
\small
{\tt 3.620}, {\tt 3.339}$_{(\times 2)}$, {\tt 3.220}, {\tt 2.974}, {\tt 2.612}, {\tt 1.797}$_{(\times 2)}$
}\end{minipage}\\
\hline
$(m^2/m_0^2)[\chi]$\quad\begin{minipage}{8.5cm}
{\vskip0.25ex
\small
{\tt 7.239}, {\tt 6.678}$_{(\times 2)}$, {\tt 6.440}, {\tt 5.948}, {\tt 5.241}$_{(\times 2)}$, {\tt 5.225}, {\tt 5.116}, {\tt 5.050}, {\tt 4.783}, {\tt 4.681}, {\tt 4.456}$_{(\times 2)}$, {\tt 4.061}$_{(\times 2)}$, {\tt 3.594}$_{(\times 2)}$, {\tt 3.095}$_{(\times 2)}$, {\tt 2.898}, {\tt 2.850}, {\tt 2.780}, {\tt 2.202}, {\tt 2.189}, {\tt 2.106}$_{(\times 2)}$, {\tt 1.852}, {\tt 1.688}, {\tt 1.551}, {\tt 1.438}$_{(\times 2)}$, {\tt 1.204}$_{(\times 2)}$, {\tt 1.142}, {\tt 1.129}, {\tt 0.927}, {\tt 0.582}, {\tt 0.310}$_{(\times 2)}$, {\tt 0.285}, {\tt 0.262}, {\tt 0.223}$_{(\times 2)}$, {\tt 0.140}, {\tt 0.089}$_{(\times 2)}$, {\tt 0.084}, {\tt 0.081}, {\tt 0.058}, {\tt 0.035}$_{(\times 2)}$, {\tt 0.029}, {\tt 0.006}, {\tt 0.002}$_{(\times 2)}$
}\end{minipage}\\
\hline
\end{longtable}

\begin{longtable}{||l||}
\hline
\raise-0.1ex\hbox{\small {\bf \#14}: $V/g^2=-14.970385$, Quality: $|Q|=10^{-6.85}$, $|\nabla|=10^{-4.10}$ $U(1)$}\\
\hline
$\phi$\quad\begin{minipage}{10.5cm}
{\vskip0.25ex
\small\tt
$+0.1728_{[1234]+}$, $+0.1197_{[1256]+}$, $+0.5219_{[1258]+}$, $-0.1197_{[1267]+}$, $-0.0480_{[1278]+}$, $-0.0480_{[1356]+}$, $+0.1197_{[1358]+}$, $-0.5219_{[1367]+}$, $+0.1197_{[1378]+}$, $+0.4398_{[1457]+}$, $-0.1728_{[1468]+}$, $+0.1728_{[2357]+}$, $-0.4398_{[2368]+}$, $-0.1197_{[2456]+}$, $+0.5219_{[2458]+}$, $-0.1197_{[2467]+}$, $+0.0480_{[2478]+}$, $-0.0480_{[3456]+}$, $-0.1197_{[3458]+}$, $+0.5219_{[3467]+}$, $+0.1197_{[3478]+}$, $+0.1728_{[5678]+}$, $+0.1364_{[1234]-}$, $+0.0967_{[1256]-}$, $-0.6919_{[1258]-}$, $-0.0967_{[1267]-}$, $+0.2809_{[1278]-}$, $+0.2809_{[1356]-}$, $+0.0967_{[1358]-}$, $+0.6919_{[1367]-}$, $+0.0967_{[1378]-}$, $-0.6119_{[1457]-}$, $-0.1364_{[1468]-}$, $-0.1364_{[2357]-}$, $-0.6119_{[2368]-}$, $+0.0967_{[2456]-}$, $+0.6919_{[2458]-}$, $+0.0967_{[2467]-}$, $+0.2809_{[2478]-}$, $-0.2809_{[3456]-}$, $+0.0967_{[3458]-}$, $+0.6919_{[3467]-}$, $-0.0967_{[3478]-}$, $-0.1364_{[5678]-}$
}\end{minipage}\\
\hline
Symmetry\quad\begin{minipage}{8.5cm}
{\vskip0.25ex
\small\tt
$-0.707\,R_{28}$ $-0.707\,R_{36}$
}\end{minipage}\\
\hline

$(m^2/m_0^2)[\psi]$\quad\begin{minipage}{8.5cm}
{\vskip0.25ex
\small
{\tt 4.064}$_{(\times 2)}$, {\tt 4.046}, {\tt 3.965}$_{(\times 2)}$, {\tt 3.626}, {\tt 2.254}, {\tt 1.568}
}\end{minipage}\\
\hline
$(m^2/m_0^2)[\chi]$\quad\begin{minipage}{8.5cm}
{\vskip0.25ex
\small
{\tt 10.570}, {\tt 10.564}, {\tt 8.127}$_{(\times 2)}$, {\tt 8.092}, {\tt 7.930}$_{(\times 2)}$, {\tt 7.251}, {\tt 5.974}$_{(\times 2)}$, {\tt 5.874}$_{(\times 2)}$, {\tt 5.766}, {\tt 5.731}, {\tt 5.469}, {\tt 5.396}, {\tt 4.546}$_{(\times 2)}$, {\tt 4.507}, {\tt 4.458}$_{(\times 2)}$, {\tt 3.661}, {\tt 3.422}, {\tt 3.136}, {\tt 2.374}$_{(\times 2)}$, {\tt 2.299}$_{(\times 2)}$, {\tt 2.020}$_{(\times 2)}$, {\tt 2.017}, {\tt 1.602}$_{(\times 2)}$, {\tt 1.594}, {\tt 1.468}$_{(\times 2)}$, {\tt 1.244}$_{(\times 2)}$, {\tt 0.983}, {\tt 0.578}, {\tt 0.206}$_{(\times 2)}$, {\tt 0.137}$_{(\times 2)}$, {\tt 0.082}, {\tt 0.072}$_{(\times 2)}$, {\tt 0.071}$_{(\times 2)}$, {\tt 0.063}$_{(\times 2)}$, {\tt 0.059}$_{(\times 2)}$, {\tt 0.039}, {\tt 0.024}, {\tt 0.022}
}\end{minipage}\\
\hline
\end{longtable}

\begin{longtable}{||l||}
\hline
\raise-0.1ex\hbox{\small {\bf \#15}: $V/g^2=-16.414456$, Quality: $|Q|=10^{-5.04}$, $|\nabla|=10^{-3.86}$ }\\
\hline
$\phi$\quad\begin{minipage}{10.5cm}
{\vskip0.25ex
\small\tt
$-0.3317_{[1234]+}$, $+0.3194_{[1236]+}$, $+0.1187_{[1245]+}$, $+0.1116_{[1256]+}$, $-0.6446_{[1278]+}$, $-0.2277_{[1347]+}$, $-0.0186_{[1358]+}$, $-0.5445_{[1367]+}$, $+0.0629_{[1457]+}$, $-0.4439_{[1468]+}$, $-0.0817_{[1567]+}$, $+0.0817_{[2348]+}$, $+0.4439_{[2357]+}$, $-0.0629_{[2368]+}$, $+0.5445_{[2458]+}$, $+0.0186_{[2467]+}$, $+0.2277_{[2568]+}$, $-0.6446_{[3456]+}$, $+0.1116_{[3478]+}$, $+0.1187_{[3678]+}$, $+0.3194_{[4578]+}$, $-0.3317_{[5678]+}$, $+0.0699_{[1234]-}$, $+0.3235_{[1236]-}$, $-0.1401_{[1245]-}$, $+0.4283_{[1256]-}$, $+0.5643_{[1278]-}$, $-0.2090_{[1347]-}$, $-0.3835_{[1358]-}$, $+0.7370_{[1367]-}$, $-0.1472_{[1457]-}$, $+0.1624_{[1468]-}$, $-0.1029_{[1567]-}$, $-0.1029_{[2348]-}$, $+0.1624_{[2357]-}$, $-0.1472_{[2368]-}$, $+0.7370_{[2458]-}$, $-0.3835_{[2467]-}$, $-0.2090_{[2568]-}$, $-0.5643_{[3456]-}$, $-0.4283_{[3478]-}$, $+0.1401_{[3678]-}$, $-0.3235_{[4578]-}$, $-0.0699_{[5678]-}$
}\end{minipage}\\
\hline
$(m^2/m_0^2)[\psi]$\quad\begin{minipage}{8.5cm}
{\vskip0.25ex
\small
{\tt 4.259}, {\tt 4.083}, {\tt 4.020}, {\tt 3.729}, {\tt 3.077}, {\tt 3.042}, {\tt 2.588}, {\tt 2.289}
}\end{minipage}\\
\hline
$(m^2/m_0^2)[\chi]$\quad\begin{minipage}{8.5cm}
{\vskip0.25ex
\small
{\tt 8.517}, {\tt 8.166}, {\tt 8.040}, {\tt 7.458}, {\tt 7.441}, {\tt 7.384}, {\tt 6.154}, {\tt 6.085}, {\tt 5.913}, {\tt 5.756}, {\tt 5.331}, {\tt 5.252}, {\tt 5.176}, {\tt 5.086}, {\tt 4.986}, {\tt 4.860}, {\tt 4.853}, {\tt 4.579}, {\tt 4.409}, {\tt 4.325}, {\tt 4.160}, {\tt 4.077}, {\tt 3.641}, {\tt 3.262}, {\tt 3.167}, {\tt 3.134}, {\tt 3.080}, {\tt 2.881}, {\tt 2.795}, {\tt 2.702}, {\tt 2.234}, {\tt 2.183}, {\tt 1.990}, {\tt 1.801}, {\tt 1.637}, {\tt 1.242}, {\tt 1.095}, {\tt 0.872}, {\tt 0.862}, {\tt 0.800}, {\tt 0.642}, {\tt 0.605}, {\tt 0.541}, {\tt 0.505}, {\tt 0.451}, {\tt 0.389}, {\tt 0.304}, {\tt 0.229}, {\tt 0.210}, {\tt 0.206}, {\tt 0.111}, {\tt 0.082}, {\tt 0.063}, {\tt 0.025}, {\tt 0.020}, {\tt 0.017}
}\end{minipage}\\
\hline
\end{longtable}

\begin{longtable}{||l||}
\hline
\raise-0.1ex\hbox{\small {\bf \#16}: $V/g^2=-17.876443$, Quality: $|Q|=10^{-2.97}$, $|\nabla|=10^{-2.26}$ }\\
\hline
$\phi$\quad\begin{minipage}{10.5cm}
{\vskip0.25ex
\small\tt
$+0.4175_{[1234]+}$, $-0.0285_{[1235]+}$, $+0.0009_{[1236]+}$, $-0.0435_{[1237]+}$, $-0.0051_{[1238]+}$, $-0.0030_{[1245]+}$, $-0.0559_{[1246]+}$, $-0.0065_{[1247]+}$, $-0.0396_{[1248]+}$, $-0.0148_{[1256]+}$, $-0.0183_{[1257]+}$, $-0.0142_{[1258]+}$, $+0.0912_{[1267]+}$, $+0.0206_{[1268]+}$, $-0.7344_{[1278]+}$, $-0.0228_{[1345]+}$, $+0.0275_{[1346]+}$, $+0.0038_{[1347]+}$, $-0.0535_{[1348]+}$, $-0.0333_{[1356]+}$, $-0.1148_{[1357]+}$, $-0.0514_{[1358]+}$, $+0.6636_{[1367]+}$, $+0.1205_{[1368]+}$, $+0.0780_{[1378]+}$, $+0.0399_{[1456]+}$, $-0.0746_{[1457]+}$, $+0.1123_{[1458]+}$, $+0.1161_{[1467]+}$, $-0.3806_{[1468]+}$, $+0.0401_{[1478]+}$, $+0.0131_{[1567]+}$, $+0.0317_{[1568]+}$, $-0.0167_{[1578]+}$, $+0.0350_{[1678]+}$, $+0.0350_{[2345]+}$, $+0.0167_{[2346]+}$, $+0.0317_{[2347]+}$, $-0.0131_{[2348]+}$, $+0.0401_{[2356]+}$, $+0.3806_{[2357]+}$, $+0.1161_{[2358]+}$, $+0.1123_{[2367]+}$, $+0.0746_{[2368]+}$, $+0.0399_{[2378]+}$, $-0.0780_{[2456]+}$, $+0.1205_{[2457]+}$, $-0.6636_{[2458]+}$, $+0.0514_{[2467]+}$, $-0.1148_{[2468]+}$, $+0.0333_{[2478]+}$, $-0.0535_{[2567]+}$, $-0.0038_{[2568]+}$, $+0.0275_{[2578]+}$, $+0.0228_{[2678]+}$, $-0.7344_{[3456]+}$, $-0.0206_{[3457]+}$, $+0.0912_{[3458]+}$, $-0.0142_{[3467]+}$, $+0.0183_{[3468]+}$, $-0.0148_{[3478]+}$, $+0.0396_{[3567]+}$, $-0.0065_{[3568]+}$, $+0.0559_{[3578]+}$, $-0.0030_{[3678]+}$, $-0.0051_{[4567]+}$, $+0.0435_{[4568]+}$, $+0.0009_{[4578]+}$, $+0.0285_{[4678]+}$, $+0.4175_{[5678]+}$, $-0.0160_{[1234]-}$, $-0.0426_{[1235]-}$, $+0.0007_{[1236]-}$, $-0.0467_{[1237]-}$, $-0.0065_{[1238]-}$, $+0.0051_{[1245]-}$, $-0.0256_{[1246]-}$, $-0.0063_{[1247]-}$, $-0.1398_{[1248]-}$, $-0.0677_{[1256]-}$, $-0.0189_{[1257]-}$, $-0.3597_{[1258]-}$, $+0.0863_{[1267]-}$, $+0.0196_{[1268]-}$, $+0.7213_{[1278]-}$, $+0.0209_{[1345]-}$, $+0.1302_{[1346]-}$, $+0.0050_{[1347]-}$, $-0.0257_{[1348]-}$, $+0.3617_{[1356]-}$, $-0.1197_{[1357]-}$, $-0.0663_{[1358]-}$, $-0.7515_{[1367]-}$, $+0.1190_{[1368]-}$, $+0.0814_{[1378]-}$, $-0.0401_{[1456]-}$, $+0.0008_{[1457]-}$, $-0.1083_{[1458]-}$, $+0.1120_{[1467]-}$, $+0.0164_{[1468]-}$, $+0.0410_{[1478]-}$, $+0.0113_{[1567]-}$, $+0.0431_{[1568]-}$, $-0.0168_{[1578]-}$, $+0.0363_{[1678]-}$, $-0.0363_{[2345]-}$, $-0.0168_{[2346]-}$, $-0.0431_{[2347]-}$, $+0.0113_{[2348]-}$, $-0.0410_{[2356]-}$, $+0.0164_{[2357]-}$, $-0.1120_{[2358]-}$, $+0.1083_{[2367]-}$, $+0.0008_{[2368]-}$, $+0.0401_{[2378]-}$, $+0.0814_{[2456]-}$, $-0.1190_{[2457]-}$, $-0.7515_{[2458]-}$, $-0.0663_{[2467]-}$, $+0.1197_{[2468]-}$, $+0.3617_{[2478]-}$, $+0.0257_{[2567]-}$, $+0.0050_{[2568]-}$, $-0.1302_{[2578]-}$, $+0.0209_{[2678]-}$, $-0.7213_{[3456]-}$, $+0.0196_{[3457]-}$, $-0.0863_{[3458]-}$, $+0.3597_{[3467]-}$, $-0.0189_{[3468]-}$, $+0.0677_{[3478]-}$, $-0.1398_{[3567]-}$, $+0.0063_{[3568]-}$, $-0.0256_{[3578]-}$, $-0.0051_{[3678]-}$, $+0.0065_{[4567]-}$, $-0.0467_{[4568]-}$, $-0.0007_{[4578]-}$, $-0.0426_{[4678]-}$, $+0.0160_{[5678]-}$
}\end{minipage}\\
\hline
$(m^2/m_0^2)[\psi]$\quad\begin{minipage}{8.5cm}
{\vskip0.25ex
\small
{\tt 4.293}$_{(\times 2)}$, {\tt 4.274}$_{(\times 2)}$, {\tt 3.027}$_{(\times 2)}$, {\tt 3.020}$_{(\times 2)}$
}\end{minipage}\\
\hline
$(m^2/m_0^2)[\chi]$\quad\begin{minipage}{8.5cm}
{\vskip0.25ex
\small
{\tt 8.585}$_{(\times 2)}$, {\tt 8.547}$_{(\times 2)}$, {\tt 7.487}$_{(\times 2)}$, {\tt 7.486}$_{(\times 2)}$, {\tt 6.055}$_{(\times 2)}$, {\tt 6.041}$_{(\times 2)}$, {\tt 5.917}$_{(\times 2)}$, {\tt 5.915}$_{(\times 2)}$, {\tt 4.952}$_{(\times 2)}$, {\tt 4.951}$_{(\times 2)}$, {\tt 4.377}$_{(\times 2)}$, {\tt 4.372}$_{(\times 2)}$, {\tt 3.665}$_{(\times 2)}$, {\tt 3.652}$_{(\times 2)}$, {\tt 2.109}$_{(\times 2)}$, {\tt 2.102}$_{(\times 2)}$, {\tt 1.559}$_{(\times 2)}$, {\tt 1.558}$_{(\times 2)}$, {\tt 1.356}$_{(\times 2)}$, {\tt 1.351}$_{(\times 2)}$, {\tt 0.818}$_{(\times 2)}$, {\tt 0.813}$_{(\times 2)}$, {\tt 0.700}$_{(\times 4)}$, {\tt 0.199}$_{(\times 4)}$, {\tt 0.029}$_{(\times 2)}$, {\tt 0.028}$_{(\times 2)}$
}\end{minipage}\\
\hline
\end{longtable}

\begin{longtable}{||l||}
\hline
\raise-0.1ex\hbox{\small {\bf \#17}: $V/g^2=-18.052693$, Quality: $|Q|=10^{-5.88}$, $|\nabla|=10^{-4.04}$ }\\
\hline
$\phi$\quad\begin{minipage}{10.5cm}
{\vskip0.25ex
\small\tt
$+0.1770_{[1235]+}$, $-0.1770_{[1236]+}$, $-0.1770_{[1245]+}$, $-0.1770_{[1246]+}$, $+0.3134_{[1278]+}$, $+0.2552_{[1348]+}$, $+0.6510_{[1357]+}$, $+0.6510_{[1367]+}$, $+0.0099_{[1457]+}$, $-0.0099_{[1467]+}$, $+0.2552_{[1568]+}$, $+0.2552_{[2347]+}$, $-0.0099_{[2358]+}$, $-0.0099_{[2368]+}$, $-0.6510_{[2458]+}$, $+0.6510_{[2468]+}$, $+0.2552_{[2567]+}$, $+0.3134_{[3456]+}$, $+0.1770_{[3578]+}$, $-0.1770_{[3678]+}$, $-0.1770_{[4578]+}$, $-0.1770_{[4678]+}$, $+0.0130_{[1235]-}$, $+0.0130_{[1236]-}$, $-0.0130_{[1245]-}$, $+0.0130_{[1246]-}$, $-0.1250_{[1348]-}$, $-0.8500_{[1357]-}$, $+0.8500_{[1367]-}$, $-0.3413_{[1457]-}$, $-0.3413_{[1467]-}$, $-0.1250_{[1568]-}$, $+0.1250_{[2347]-}$, $+0.3413_{[2358]-}$, $-0.3413_{[2368]-}$, $+0.8500_{[2458]-}$, $+0.8500_{[2468]-}$, $+0.1250_{[2567]-}$, $+0.0130_{[3578]-}$, $+0.0130_{[3678]-}$, $-0.0130_{[4578]-}$, $+0.0130_{[4678]-}$
}\end{minipage}\\
\hline
$(m^2/m_0^2)[\psi]$\quad\begin{minipage}{8.5cm}
{\vskip0.25ex
\small
{\tt 4.398}$_{(\times 2)}$, {\tt 3.811}$_{(\times 2)}$, {\tt 3.346}$_{(\times 2)}$, {\tt 2.359}$_{(\times 2)}$
}\end{minipage}\\
\hline
$(m^2/m_0^2)[\chi]$\quad\begin{minipage}{8.5cm}
{\vskip0.25ex
\small
{\tt 8.796}$_{(\times 2)}$, {\tt 7.623}$_{(\times 2)}$, {\tt 6.923}$_{(\times 2)}$, {\tt 6.900}$_{(\times 2)}$, {\tt 6.692}$_{(\times 2)}$, {\tt 5.596}$_{(\times 2)}$, {\tt 4.874}$_{(\times 2)}$, {\tt 4.719}$_{(\times 2)}$, {\tt 4.279}$_{(\times 2)}$, {\tt 4.258}$_{(\times 2)}$, {\tt 4.176}$_{(\times 2)}$, {\tt 4.036}$_{(\times 2)}$, {\tt 3.482}$_{(\times 2)}$, {\tt 3.200}$_{(\times 2)}$, {\tt 2.963}$_{(\times 2)}$, {\tt 2.305}$_{(\times 2)}$, {\tt 2.010}$_{(\times 2)}$, {\tt 1.522}$_{(\times 2)}$, {\tt 1.099}$_{(\times 2)}$, {\tt 1.067}$_{(\times 2)}$, {\tt 1.060}$_{(\times 2)}$, {\tt 0.924}$_{(\times 2)}$, {\tt 0.440}$_{(\times 2)}$, {\tt 0.126}$_{(\times 2)}$, {\tt 0.070}$_{(\times 2)}$, {\tt 0.043}$_{(\times 2)}$, {\tt 0.036}$_{(\times 2)}$, {\tt 0.016}$_{(\times 2)}$
}\end{minipage}\\
\hline
\end{longtable}

\begin{longtable}{||l||}
\hline
\raise-0.1ex\hbox{\small {\bf \#18}: $V/g^2=-21.265976$, Quality: $|Q|=10^{-6.16}$, $|\nabla|=10^{-3.67}$ }\\
\hline
$\phi$\quad\begin{minipage}{10.5cm}
{\vskip0.25ex
\small\tt
$-0.5622_{[1234]+}$, $+0.2761_{[1248]+}$, $-0.0763_{[1256]+}$, $+0.3668_{[1257]+}$, $-0.0763_{[1268]+}$, $-0.3668_{[1278]+}$, $+0.0447_{[1358]+}$, $+0.7687_{[1367]+}$, $-0.3668_{[1456]+}$, $-0.0763_{[1457]+}$, $+0.3668_{[1468]+}$, $-0.0763_{[1478]+}$, $+0.2761_{[1678]+}$, $+0.2761_{[2345]+}$, $-0.0763_{[2356]+}$, $-0.3668_{[2357]+}$, $+0.0763_{[2368]+}$, $-0.3668_{[2378]+}$, $-0.7687_{[2458]+}$, $-0.0447_{[2467]+}$, $-0.3668_{[3456]+}$, $+0.0763_{[3457]+}$, $-0.3668_{[3468]+}$, $-0.0763_{[3478]+}$, $-0.2761_{[3567]+}$, $-0.5622_{[5678]+}$, $-0.4045_{[1234]-}$, $-0.2766_{[1248]-}$, $+0.3801_{[1257]-}$, $+0.3801_{[1278]-}$, $-0.9522_{[1367]-}$, $+0.3801_{[1456]-}$, $+0.3801_{[1468]-}$, $+0.2766_{[1678]-}$, $-0.2766_{[2345]-}$, $+0.3801_{[2357]-}$, $-0.3801_{[2378]-}$, $-0.9522_{[2458]-}$, $-0.3801_{[3456]-}$, $+0.3801_{[3468]-}$, $-0.2766_{[3567]-}$, $+0.4045_{[5678]-}$
}\end{minipage}\\
\hline
$(m^2/m_0^2)[\psi]$\quad\begin{minipage}{8.5cm}
{\vskip0.25ex
\small
{\tt 5.785}$_{(\times 2)}$, {\tt 4.194}$_{(\times 4)}$, {\tt 3.697}$_{(\times 2)}$
}\end{minipage}\\
\hline
$(m^2/m_0^2)[\chi]$\quad\begin{minipage}{8.5cm}
{\vskip0.25ex
\small
{\tt 11.570}$_{(\times 2)}$, {\tt 9.001}$_{(\times 2)}$, {\tt 8.471}$_{(\times 4)}$, {\tt 8.388}$_{(\times 4)}$, {\tt 7.554}$_{(\times 2)}$, {\tt 7.394}$_{(\times 2)}$, {\tt 7.065}$_{(\times 2)}$, {\tt 6.352}$_{(\times 2)}$, {\tt 6.220}$_{(\times 4)}$, {\tt 4.597}$_{(\times 2)}$, {\tt 4.407}$_{(\times 4)}$, {\tt 2.706}$_{(\times 2)}$, {\tt 2.444}$_{(\times 2)}$, {\tt 1.853}$_{(\times 4)}$, {\tt 1.476}$_{(\times 2)}$, {\tt 1.331}$_{(\times 2)}$, {\tt 1.303}$_{(\times 4)}$, {\tt 0.693}$_{(\times 2)}$, {\tt 0.569}$_{(\times 2)}$, {\tt 0.293}$_{(\times 4)}$, {\tt 0.210}$_{(\times 2)}$
}\end{minipage}\\
\hline
\end{longtable}

\begin{longtable}{||l||}
\hline
\raise-0.1ex\hbox{\small {\bf \#19}: $V/g^2=-21.408498$, Quality: $|Q|=10^{-3.63}$, $|\nabla|=10^{-2.84}$ }\\
\hline
$\phi$\quad\begin{minipage}{10.5cm}
{\vskip0.25ex
\small\tt
$+0.2178_{[1234]+}$, $-0.6540_{[1256]+}$, $-0.0068_{[1267]+}$, $+0.6212_{[1278]+}$, $-0.0068_{[1356]+}$, $-0.1884_{[1358]+}$, $-0.6540_{[1367]+}$, $+0.2178_{[1457]+}$, $-0.2704_{[1468]+}$, $+0.2704_{[2357]+}$, $-0.2178_{[2368]+}$, $+0.6540_{[2458]+}$, $+0.1884_{[2467]+}$, $+0.0068_{[2478]+}$, $+0.6212_{[3456]+}$, $-0.0068_{[3458]+}$, $-0.6540_{[3478]+}$, $+0.2178_{[5678]+}$, $+0.3332_{[1234]-}$, $+0.9511_{[1256]-}$, $+0.0004_{[1258]-}$, $-0.0055_{[1267]-}$, $-0.0055_{[1356]-}$, $+0.9511_{[1367]-}$, $-0.0004_{[1378]-}$, $-0.3332_{[1457]-}$, $-0.3332_{[2368]-}$, $-0.0004_{[2456]-}$, $+0.9511_{[2458]-}$, $-0.0055_{[2478]-}$, $+0.0055_{[3458]-}$, $-0.0004_{[3467]-}$, $-0.9511_{[3478]-}$, $-0.3332_{[5678]-}$
}\end{minipage}\\
\hline
$(m^2/m_0^2)[\psi]$\quad\begin{minipage}{8.5cm}
{\vskip0.25ex
\small
{\tt 5.055}$_{(\times 2)}$, {\tt 5.052}$_{(\times 2)}$, {\tt 3.957}$_{(\times 4)}$
}\end{minipage}\\
\hline
$(m^2/m_0^2)[\chi]$\quad\begin{minipage}{8.5cm}
{\vskip0.25ex
\small
{\tt 11.250}$_{(\times 4)}$, {\tt 10.109}$_{(\times 2)}$, {\tt 10.105}$_{(\times 2)}$, {\tt 7.914}$_{(\times 4)}$, {\tt 7.857}$_{(\times 4)}$, {\tt 5.251}$_{(\times 4)}$, {\tt 4.998}$_{(\times 4)}$, {\tt 4.973}$_{(\times 2)}$, {\tt 4.965}$_{(\times 2)}$, {\tt 4.459}$_{(\times 4)}$, {\tt 2.159}$_{(\times 2)}$, {\tt 2.153}$_{(\times 2)}$, {\tt 1.577}$_{(\times 2)}$, {\tt 1.574}$_{(\times 2)}$, {\tt 1.273}$_{(\times 4)}$, {\tt 0.883}$_{(\times 4)}$, {\tt 0.222}$_{(\times 4)}$, {\tt 0.181}$_{(\times 2)}$, {\tt 0.179}$_{(\times 2)}$
}\end{minipage}\\
\hline
\end{longtable}

\begin{longtable}{||l||}
\hline
\raise-0.1ex\hbox{\small {\bf \#20}: $V/g^2=-25.149369$, Quality: $|Q|=10^{-6.30}$, $|\nabla|=10^{-3.50}$ }\\
\hline
$\phi$\quad\begin{minipage}{10.5cm}
{\vskip0.25ex
\small\tt
$-0.1859_{[1234]+}$, $+0.3286_{[1256]+}$, $+0.0475_{[1257]+}$, $-0.0475_{[1268]+}$, $-0.3287_{[1278]+}$, $-0.0852_{[1356]+}$, $-0.3827_{[1357]+}$, $+0.0062_{[1358]+}$, $-0.2180_{[1367]+}$, $-0.3826_{[1368]+}$, $-0.0851_{[1378]+}$, $-0.1129_{[1456]+}$, $+0.1179_{[1457]+}$, $-1.0129_{[1458]+}$, $-0.2124_{[1467]+}$, $+0.1181_{[1468]+}$, $-0.1128_{[1478]+}$, $-0.1128_{[2356]+}$, $-0.1181_{[2357]+}$, $-0.2124_{[2358]+}$, $-1.0129_{[2367]+}$, $-0.1179_{[2368]+}$, $-0.1129_{[2378]+}$, $+0.0851_{[2456]+}$, $-0.3826_{[2457]+}$, $+0.2180_{[2458]+}$, $-0.0062_{[2467]+}$, $-0.3827_{[2468]+}$, $+0.0852_{[2478]+}$, $-0.3287_{[3456]+}$, $+0.0475_{[3457]+}$, $-0.0475_{[3468]+}$, $+0.3286_{[3478]+}$, $-0.1859_{[5678]+}$, $+0.1870_{[1234]-}$, $-0.0641_{[1256]-}$, $-0.4430_{[1257]-}$, $+0.4430_{[1268]-}$, $+0.0642_{[1278]-}$, $-0.0403_{[1356]-}$, $+0.7939_{[1357]-}$, $+0.2410_{[1358]-}$, $+0.2494_{[1367]-}$, $+0.7937_{[1368]-}$, $-0.0402_{[1378]-}$, $-0.4082_{[1456]-}$, $+0.0520_{[1457]-}$, $-0.2165_{[1458]-}$, $-0.0740_{[1467]-}$, $+0.0520_{[1468]-}$, $-0.4082_{[1478]-}$, $+0.4082_{[2356]-}$, $+0.0520_{[2357]-}$, $+0.0740_{[2358]-}$, $+0.2165_{[2367]-}$, $+0.0520_{[2368]-}$, $+0.4082_{[2378]-}$, $-0.0402_{[2456]-}$, $-0.7937_{[2457]-}$, $+0.2494_{[2458]-}$, $+0.2410_{[2467]-}$, $-0.7939_{[2468]-}$, $-0.0403_{[2478]-}$, $-0.0642_{[3456]-}$, $+0.4430_{[3457]-}$, $-0.4430_{[3468]-}$, $+0.0641_{[3478]-}$, $-0.1870_{[5678]-}$
}\end{minipage}\\
\hline
$(m^2/m_0^2)[\psi]$\quad\begin{minipage}{8.5cm}
{\vskip0.25ex
\small
{\tt 7.557}, {\tt 6.906}, {\tt 5.728}, {\tt 5.212}, {\tt 4.283}, {\tt 3.325}, {\tt 3.265}, {\tt 2.715}
}\end{minipage}\\
\hline
$(m^2/m_0^2)[\chi]$\quad\begin{minipage}{8.5cm}
{\vskip0.25ex
\small
{\tt 15.114}, {\tt 13.813}, {\tt 13.636}, {\tt 13.631}, {\tt 11.455}, {\tt 10.424}, {\tt 9.730}, {\tt 9.728}, {\tt 9.169}, {\tt 9.163}, {\tt 8.566}, {\tt 7.200}, {\tt 7.106}, {\tt 7.010}, {\tt 6.888}, {\tt 6.866}, {\tt 6.761}, {\tt 6.650}, {\tt 6.530}, {\tt 6.016}, {\tt 6.015}, {\tt 5.609}, {\tt 5.432}, {\tt 5.430}, {\tt 5.379}, {\tt 5.311}, {\tt 5.268}, {\tt 5.058}, {\tt 4.398}, {\tt 4.379}, {\tt 4.106}, {\tt 3.962}, {\tt 3.869}, {\tt 3.761}, {\tt 3.544}, {\tt 3.367}, {\tt 2.365}, {\tt 2.226}, {\tt 2.181}, {\tt 2.163}, {\tt 1.437}, {\tt 1.333}, {\tt 1.232}, {\tt 1.221}, {\tt 0.952}, {\tt 0.844}, {\tt 0.816}, {\tt 0.580}, {\tt 0.376}, {\tt 0.315}, {\tt 0.237}, {\tt 0.219}, {\tt 0.053}, {\tt 0.021}, {\tt 0.002}, {\tt 0.001}
}\end{minipage}\\
\hline
\end{longtable}

\end{document}